\documentclass[10pt,journal,final,letterpaper,twocolumn,twoside]{IEEEtran}

\usepackage{amsmath,amsfonts,bm,amssymb,psfrag,ifthen,color,subfigure}
\usepackage{mathrsfs}
\usepackage[justification=centering,font=footnotesize]{caption}
\usepackage{todonotes}

\allowdisplaybreaks[4]

\usepackage[dvips]{epsfig}
\usepackage[dvips]{graphicx}
\usepackage{amsmath,amsfonts,bm,amssymb,psfrag,ifthen,color,subfigure}
\usepackage{algpseudocode}
\usepackage{algorithm}
\usepackage{algorithmicx}
\usepackage{ifthen}
\usepackage{setspace}
\usepackage{cite}
\usepackage{url}
\usepackage{longtable}
\usepackage{stfloats}  
\usepackage{graphics,booktabs,color,subfigure}
\usepackage{float}
\usepackage{diagbox} 

\newcommand{\tabincell}[2]{\begin{tabular}{@{}#1@{}}#2\end{tabular}}

\usepackage{listings}

\floatname{algorithm}{Algorithm}

\newcounter{mytempeqncnt}

\begin{document}
\title{Spectral  and Energy Efficiency of DCO-OFDM in Visible
Light Communication Systems with  Finite-Alphabet Inputs}
  \author{Ruixin Yang, Shuai Ma, Zihan Xu,  Hang Li, Xiaodong Liu, Xintong Ling, Xiong Deng, Xun Zhang, Shiyin Li
\thanks{R. Yang, S. Ma and S. Li   are with the School of Information and Control   Engineering, China
University of Mining and Technology, Xuzhou, 221116,
China. (e-mail: \{ray.young, mashuai001, lishiyin\}@cumt.edu.cn).}
\thanks{Z. Xu is with Spreadtrum Communications (Shanghai) Co., Ltd., Shanghai 201203, China (e-mail: zihan.xu@unisoc.com).}
\thanks{H. Li is with the Shenzhen Research Institute of Big Data, Shenzhen 518172, Guangdong, China. (email: hangdavidli@163.com).}
 \thanks{X. Liu is with the School of Information Engineering, Nanchang University, Nanchang 330031, China (e-mail: xiaodongliu@ieee.org). }
 \thanks{X. Ling is with the National Mobile Communications Research Laboratory, Southeast University,
 and the Purple Mountain Laboratories, Nanjing, China. (e-mail: xtling@seu.edu.cn). }
  \thanks{X. Deng is with the Center for Information Photonics and Communications, Southwest Jiaotong University, Chengdu 610031, China.(email: xiongdeng@swjtu.edu.cn).}
 \thanks{X. Zhang is with Institut Suprieur dElectronique de Paris, ISEP Paris, France. (e-mail: xun.zhang@isep.fr).}

}

\maketitle
\begin{abstract}
The
{bound of the information transmission rate}
 of direct current biased optical orthogonal frequency division multiplexing (DCO-OFDM)
for visible light communication (VLC) with finite-alphabet inputs is yet unknown,
{where the corresponding spectral efficiency (SE) and energy efficiency (EE) stems out as
the open research problems.}
In this paper, we  derive the exact achievable rate of {the} DCO-OFDM system
with finite-alphabet inputs for the first time.
Furthermore, we investigate   SE  maximization problems of {the} DCO-OFDM system
subject to  both electrical  and optical power constraints.
By exploiting the relationship between the mutual information and the minimum mean-squared error,
we propose a multi-level mercury-water-filling  power allocation scheme
to achieve the  maximum  SE.
Moreover,    the    EE maximization problems of {the} DCO-OFDM system are studied, and  the Dinkelbach-type power allocation scheme is    developed for  the  maximum  EE.
 Numerical results  verify    the   effectiveness of the proposed theories and power allocation schemes.

\end{abstract}

\IEEEpeerreviewmaketitle

\section{Introduction}

\subsection{Motivation and Contributions}

As the number of
Internet of Things (IoT) devices increases tremendously,
the radio frequency (RF) wireless networks are facing  {an} ever-growing bandwidth burden to
support huge and high-speed data transfer \cite{Razzaque_IoT_2016,Zanella_IoT_2014,Wong_2017_5G}.
It is reported that more than  $80\% $ of wireless data is generated in the indoor environment \cite{Feng_WC_2018,Diamantoulakis_TGCN_2018}.
As a result, visible light communication (VLC)\cite{IEEE_Std}
has emerged as a promising technology to provide {high-speed} data transmission and illumination service
simultaneously due to the huge unlicensed visible light spectrum for future  IoT  applications \cite{Hass_JLT_2016,Liu_TC_2020}. 
{By using the ordinary light emitting diodes (LEDs) at the transmitter side and the simple intensity modulation and direct detection (IM/DD) at the receiver side,}
 VLC can simultaneously support high-speed communication and  illumination
 without  electromagnetic interference to the conventional RF networks \cite{Jovicic_CM_2013,Zhang_WC_2015,Pathak_CST_2015,Elgala_CM_2011}.

From the wireless communication point of view, the VLC system, similar {to the RF system}, still faces the issue {of inter-symbol interference} (ISI) {resulting} from {limited} modulation bandwidth  of {LEDs} \cite{Le_Minh_PTL_2008} and multipath distortion \cite{Hussein_JLT_2015}.
Thus, the orthogonal frequency division multiplexing (OFDM) based techniques
are introduced  into the VLC system   against the ISI and enhance the communication capacity.
There are two typical OFDM-based transmission schemes:
direct current (DC) biased optical OFDM (DCO-OFDM)
and asymmetrically clipped optical OFDM (ACO-OFDM).
Specifically, to guarantee that the transmitted signals are  positive,  DCO-OFDM adds a DC-bias to the time-domain signals and clips the
remaining negative signals to zero, while ACO-OFDM only transmits the positive parts of the OFDM waveform.
Note that if the DC-bias power is neglected,  DCO-OFDM can achieve the Shannon {capacity while} ACO-OFDM has a 3-dB penalty \cite{Dimitrov_JLT_2013}.
In practice, the DCO-OFDM system usually operates with finite-alphabet inputs such as pulse amplitude modulation (PAM)
and quadrature amplitude modulation (QAM).

Most of the existing {literature} on the DCO-OFDM system adopted the assumption that
if the number of subcarriers is {large,} the time domain signal after the inverse fast Fourier
transform (IFFT) is approximately Gaussian distributed \cite{Wu_JLT_2015,Deng_Access_2019,Ling_TSP_2016_offset,Ling_JSAC_2018_biased}.
{However, according to the central limit theorem, only if the number of subcarriers tends to infinity, and the frequency domain symbols are independent, the time domain signal equivalently follows the Gaussian distribution, whatever the alphabet set is a continuous set or a finite constellation set. Therefore, some approximation errors exist to some extent. Meanwhile, when the random process is unbounded, the clipping operator cannot be avoided, and  the clipping noise may cause the inevitable information loss\cite{Armstrong_CL_2008,Ling_TSP_2016_offset}.}
Under such {an} assumption, existing models cannot accurately depict the {achievable rate} of the DCO-OFDM system with finite-alphabet inputs,
and the corresponding 
{bound of the information transmission rate}
is yet unknown.
Thus, the spectrum efficiency  (SE) and  energy efficiency  (EE) of the DCO-OFDM system still need further investigation.

Besides, unlike the conventional RF communication systems, where only the electrical power constraint is considered in most scenarios,
the average optical power constraint plays an important role in the illumination requirement \cite{Jovicic_CM_2013,Pathak_CST_2015}
and should also be considered in the discussions of SE and EE.

With the aforementioned issues, in this paper, we consider such a typical DCO-OFDM-based VLC system
and investigate the optimal power allocation over subcarriers
to maximize the SE  and  EE.  
Our main contributions of this paper are summarized as follows:

\begin{itemize}
\item
To our best knowledge, the 
{theoretical bound of the information transmission rate}
of the  DCO-OFDM system with finite-alphabet inputs remains unknown.
In this work, we derive the information  rate of the DCO-OFDM system  with finite-alphabet inputs for the first time.
Specifically, 
we derive the exact achievable rate expression of the DCO-OFDM system without the information loss.
Moreover, we derive  the closed-form lower  bound for the derived achievable rate.
The obtained expressions can be used as the performance metrics{,} and we also apply them in the transmission design.

\item
Based on the closed-form lower bound of the achievable rate,
we jointly optimize DC-bias{,} and power allocation  of subcarriers to maximize the SE of the DCO-OFDM system
under  both average optical power and total electrical {transmitted} power constraints.
We find that the optimal  DC-bias without information loss can be presented in a closed-form expression.
Then, by restricting both optical and electrical power constraints, the optimal power allocation can be obtained by employing the interior-point algorithm.

\item
Next, we further study the SE maximization problem based on the exact achievable rate expression under the same constraints above.
By exploiting the  Karush-Kuhn-Tucker (KKT) conditions
and the relationship between the mutual information and the minimum mean-squared error (MMSE)\footnote{$\frac{\partial }{{\partial {\rm{SNR}}}}I\left( {{\rm{SNR}}} \right) = {\rm{MMSE}}\left( {{\rm{SNR}}} \right)$,
where SNR denotes  signal-to-noise ratio.}\cite{Guo_TIT_2005},
we propose a multi-level mercury-water-filling  power allocation scheme  to achieve the  maximum of SE {\cite{Zhang_2008_JSTSP, Lozano_TIT_2006}}. It is shown that the power allocation behaves differently with respect to the channel gain in the low {and high} power domains.

\item  Finally,
we investigate the power allocation to maximize the EE of the DCO-OFDM system derived {two achievable rate metrics}, respectively.
We employ {the} Dinkelbach-type algorithm  to convert the concave-linear fractional problem into a sequence of convex sub-problems,
and then obtain the optimal power allocation by the interior-point algorithm.
In addition, we reveal that the optimal power allocation of the EE maximization problem with finite-alphabet inputs is related to the SE requirement.
For the low  SE requirement, the  allocated power of each subcarrier is proportional to the channel gain.
While for the high SE requirement, the allocated power of  each  subcarrier is inversely proportional to the channel gain.


\end{itemize}

\subsection{Related Works and Organization}



Most of the existing {literature} on SE and EE of {the} DCO-OFDM system study that
the time domain signal output by IFFT is approximately Gaussian distributed when the number of subcarriers is large  \cite{Wu_JLT_2015,Deng_Access_2019,Ling_TSP_2016_offset,Ling_JSAC_2018_biased},
which causes signal approximation errors, as well as clipping noise and information loss with the clipping process \cite{Armstrong_CL_2008,Ling_TSP_2016_offset}.

It should be pointed out that the SE of the DCO-OFDM system has been extensively studied for different input  constraints.
For example, under the  optical power  constraint and a target bit error rate (BER),
the  adaptive modulation scheme was employed in \cite{Wu_JLT_2015} to maximize the SE,
and it {was shown} that the SE of ACO-OFDM is higher than that of DCO-OFDM in the low-SNR region
while it has a $50\% $ reduction compared to DCO-OFDM at high-SNR region.
For uniform power within an optimized band, the authors in \cite{Mardanikorani_TC_2020}  proposed a  discrete bit loading algorithm
to maximize the achievable rate of the DCO-OFDM system. 
For the single-LED case,
the signal-to-noise-plus-distortion ratio (SNDR) of DCO-OFDM was maximized by
jointly  optimizing  both  the DC-bias and  the information-carrying
power under both the optical and  electrical power constraints \cite{Ling_TSP_2016_offset}.
In the case with multiple LEDs,
the SNDR maximization problem in the DCO-OFDM system was studied in \cite{Ling_JSAC_2018_biased}
by properly designing the biased beamforming with the optical power constraint.
By utilizing generalized mutual information, the lower bound of information rate was
derived in \cite{ZhouJing_TC_2018} for the DCO-OFDM system under the average optical power constraint.

Recently, due to the increasing number of IoT devices,
the research on EE has attracted great {attention} to reduce {consumption} and prolong the lifetime \cite{Kashef_JSAC_2016,Guo_IoT_2021,Xiong_TWC_2011}.
It was reported in \cite{Deng_Access_2019} that both the energy and spectrum efficiency achieved with DCO-OFDM is higher than that obtained by ACO-OFDM in the case that a constant DC-bias power is given for the  illumination requirement.
 By replacing the negative parts of signals with their absolute values,
a power-efficient symbol recovery scheme  for {the} DCO-OFDM system
 was proposed in \cite{Weiss_JLT_2016}, which can improve the symbol error rate (SER) performance for a given DC-bias.
 Under the constraints of the BER and the total transmitted power,
 the authors in \cite{Hei_TVT_2019} investigated the achievable rate maximization problem for the  DCO-OFDM system,
  and further optimized transmitted power to achieve the tradeoff between SE and EE.

It was reported in \cite{Xiao_TSP_2011} that the mutual information maximizing design and classic power allocation scheme
based on Gaussian distributed assumption will lead to {a} significant loss.
However, to the best of our knowledge, only {a} few works have investigated the SE and EE of {the} DCO-OFDM with finite-alphabet inputs.
The most relevant work is \cite{Ge_WCSP_2017}, where
the EE maximization problem of DCO-OFDM was studied
by designing the optimal solution of the modulation order and power allocation
under the minimum SE requirement and {a} total {transmitted power} constraint.
However, since the mutual information does not have the closed-form expression,
the authors in \cite{Ge_WCSP_2017} adopted the closed-form lower bound and upper bound  to approximate
the exact mutual information to solve the optimal power allocation problem.
In our study, we exploit the relationship between the mutual information and MMSE
to deal with the non-closed-form expression.
Moreover, we study the 
{bound of the information transmission rate}
of the DCO-OFDM system {based on finite-alphabet inputs and the optimal DC-bias such that the clipping can be avoided},
which is different from \cite{Ling_TSP_2016_offset} and \cite{Ling_JSAC_2018_biased} with clipping noise,
and propose the optimal power allocation scheme to
maximize SE and EE of the DCO-OFDM system
under the constraints of average optical power and total electrical {transmitted} power,
and further consider the minimum SE requirement when maximizing EE.

The rest of this paper is organized as follows.
The system model of the considered DCO-OFDM  system is presented in Section II.
The achievable rate expressions of  DCO-OFDM  are derived in Section III.
The SE and EE maximization problems of the DCO-OFDM system are respectively studied in Section IV and Section V.
The simulation results are presented in Section VI.
Finally, the conclusions are drawn in Section VII.

\emph{Notations}: 
 Expected value of  a random variable $z$ is denoted by ${\mathbb{E}}\left\{ z \right\}$. ${\left(  \cdot  \right)^ * }$ represents conjugate transformation.
 ${\left[  x  \right]^ + }$ denotes $\max \left\{ {x,0} \right\}$.
 ${\mathop{\rm Re}\nolimits} \left(  \cdot  \right)$ and ${\mathop{\rm Im}\nolimits} \left(  \cdot  \right)$
denote the real and imaginary parts of their argument, respectively.
${{\partial f\left( \cdot \right)} \mathord{\left/
 {\vphantom {{\partial f\left( \cdot \right)} {\partial x}}} \right.
 \kern-\nulldelimiterspace} {\partial x}}$ represents the partial derivative of function ${f\left(  \cdot  \right)}$.
Given a  variable $y$, ${\mathbb{E}}\left\{ {z\left| y \right.} \right\}$ represents the
conditional expectation of $z$ for given $y$.
$I\left( {X;Y} \right)$ represents the mutual information of $X$ and $Y$.


\section{System Model}

 \begin{figure}[htbp]
          \centering
      \includegraphics[width =.45\textwidth]{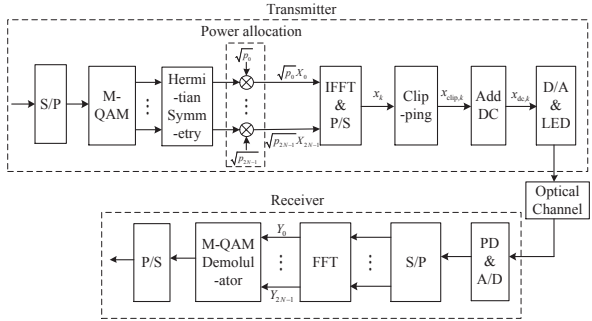}
 \caption{~The {schematic diagram} of a DCO-OFDM VLC system.}
  \label{system}
\end{figure}
In this paper, we consider both the SE and EE optimization problems
for the DCO-OFDM system with  finite-alphabet inputs  shown in Fig. 1.
 At the transmitter side, the input bit streams are modulated by an $M$-ary
QAM after serial-to-parallel (S/P) conversion.
{Due to the IM/DD, the transmit signal is required not only  to be non-negative but also to be real-valued.}
In order to ensure that the output signal of the IFFT is a real-valued VLC signal,
the IFFT input symbols of $2N$ subcarriers after Hermitian symmetry should satisfy
  \begin{align}\label{Herm}
& \sqrt {{p_{2N-i}}}{X_{2N - i}} = \sqrt{p_{i}}X_i^*,\quad i = 1,...,N - 1,
 \end{align}
{where ${X_{i}}$ is  the  signal of the $i$th subcarrier, which should be $ {X_0}$ = ${X_N}$ = 0 to block the DC component. Without loss of generality, we can assume $\mathbb{E}\left\{ {{{\left| {{X_i}} \right|}^2}} \right\} = 1$. Besides, let $p_i$ denote the allocated power for the $i$th subcarrier, and follow $p_0=p_N=0$, and $p_{2N-i}=p_i, i=1,\dots,N-1$.}
 Then, the IFFT output signal ${x_k}$ in the time domanoin is given by
\begin{subequations}
 \begin{align}
 {x_k} = & \frac{1}{{\sqrt {2N} }}\sum\limits_{i = 0}^{2N - 1} {\sqrt {{p_i}} {X_i}\exp \left( {j\frac{{\pi ki}}{N}} \right)}  \\
    = & \frac{1}{{\sqrt {2N} }}\left(\sum\limits_{i = 1}^{N - 1} \sqrt {{p_i}} {X_i}\exp \left( {j\frac{{\pi ki}}{N}} \right)  \right.\nonumber\\
        &~~~~~~~~~~~~~~~~~\left.+ \sum\limits_{i = 1}^{N - 1} {\left( {\sqrt {{p_i}} {X_i}\exp \left( {j\frac{{\pi ki}}{N}} \right)} \right)}^*  \right)  \\
   =&\sqrt {\frac{2}{N}} \sum\limits_{i = 1}^{N - 1} {\sqrt {{p_i}} {\rm{Re}}\left( {{X_i}\exp \left( {j\frac{{\pi ki}}{N}} \right)} \right)},  \label{IFFT_1}
 \end{align}
 \end{subequations}
 where $k = 0,...,2N-1$.  It is easy to find that $\mathbb{E}\left\{ {x_{k}} \right\} = 0$.


To guarantee the  VLC signal non-negative, the clipping operator is applied.
Specifically, the time domain signal ${x_k}$ is converted to clipped signal ${x_{{\rm{clip}},k}} $
by clipping at the level of $-{I_{{\rm{dc}}}}$, where ${I_{{\rm{dc}}}}$ is the DC-bias and the clipping operator is defined as
\begin{align}
{x_{{\rm{clip}},k}}  = \left\{ {\begin{array}{*{20}{c}}
    x_k \\
   - {I_{{\rm{dc}}}}  \\
\end{array}} \right.\begin{array}{*{20}{c}}
   {x_k} \ge  - {I_{{\rm{dc}}}},  \\
   {x_k} \le  - {I_{{\rm{dc}}}}.  \\
\end{array}
\end{align}
Then ${x_{{\rm{clip}},k}}$  is added with $I_{\rm{dc}}$ which only affects the $0$th subcarrier in the frequency domain.
Thus, we obtain the non-negative signal
\begin{align}   \label{x_dc}
{x_{{\rm{dc}},k}} = {x_{{\rm{clip}},k}} + {I_{{\rm{dc}}}}.
\end{align}

{
  If we want to avoid the information loss brought by the clipping noise, a feasible way is to avoid the clipping operation, i.e., the amplitude of the time domain signal $x_k$ should be bounded, and an appropriate DC-bias $I_{\mathrm{dc}}$ should satisfy
     \begin{align}  \label{x_k_idc_org}
      {x_k} + {I_{{\rm{dc}}}} \geq \min\left\{x_k\right\} + {I_{{\rm{dc}}}} \geq 0.
      \end{align}
 }
Moreover, the digital signal ${x_{{\rm{dc}},k}}$ is converted to the analog signal
via digital-to-analog convertor (DAC) and then   transmitted   by LED.
To satisfy the illumination and human eye safety requirements,
the average optical power is restricted, i.e.,
 \begin{align}\label{dimming_aco_org}
\mathbb{E}\left\{ {x_{{\rm{dc}},k}} \right\} \le  {P_o},
\end{align}
where  ${P_o}$ represents the maximum average optical power budget.

Besides, for the practical electrical circuits consideration,
the total electrical {transmitted} power  of the  VLC system should also be constrained, i.e.,
\begin{align} \label{electrical_aco_org}
\sum\limits_{k = 0}^{2N - 1} {\mathbb{E}\left\{ {x_{{\rm{dc}},k}^2} \right\}}  \le {P_e}.
 \end{align}
 where $P_e$ is the maximum total electrical {transmitted} power budget.

 Note that the  channel of {the} VLC  system is generally lowpass \cite{Kahn_1997,Ling_TSP_2016_offset}, 
and the channel gain difference between each subcarrier  can be utilized
via allocating proper power to each subcarrier to improve the performance of the VLC system.
Indoor VLC channel includes two components:
the line-of-sight (LOS) link between the transmitter and the receiver;
and the diffuse link that is the superposition of all non-LOS components caused by one or more reflections on the surface of the room.

Let ${H_{{\rm{L}},i}} = {\eta _{\rm{L}}}{e^{ - j2\pi {f_i}{\tau _{\rm{L}}}}}$ denote the  channel gain of the LOS link of the $i$th subcarrier,
where  ${\eta _{\rm{L}}}$ is the  generalized Lambertian radiator \cite{Liu_TC_2020,Kahn_1997} expressed as
\begin{align}\label{Lambertian}
{\eta _{\rm{L}}} = \frac{{\left( {m + 1} \right){A_{\rm{r}}}\cos \left( \varphi  \right)}}{{2\pi {d^2}}}{\cos ^m}\left( \theta  \right)T\left( \varphi  \right)G\left( \varphi  \right){\rm{rect}}\left( {{\varphi  \mathord{\left/
 {\vphantom {\varphi  \Psi }} \right.
 \kern-\nulldelimiterspace} \Psi }} \right).
\end{align}
In \eqref{Lambertian}, $m =  - \ln 2/\ln \left( {\cos {\Phi _{1/2}}} \right)$ denotes the order of Lambertian emission,
 and ${{\Phi _{1/2}}}$ is the semi-angle at half power.
${A_{\rm{r}}}$ denotes the effective detector area of the {photodetector} (PD),
$\varphi$  and $\theta$ are the incidence and irradiance angles from the LED to the PD, respectively,
${T\left( \varphi  \right)}$ and ${G\left( \varphi  \right)}$ are  the optical filter gain and the concentrator gain of the receiver, respectively,
$\Psi $ represents the field-of-view (FOV) of the receiver,
the rectangular function ${\rm{rect}}\left( x \right)$ takes 1 whenever $\left| x \right| \le 1$, and is 0 otherwise,
 $f_i$ denotes  the frequency of the $i$th subcarrier,
 $\tau _{\rm{L}}  = d/c$ is the signal propagation delay of the LOS link between the LED to the PD,
  ${{d}}$ is the distance between the LED to the PD,
 and $c$ stands for the speed of light.

Let ${H_{{\rm{D}},i}} = \frac{{{\eta _{\rm{D}}}}}{{1 + j2\pi {\tau _{\rm{D}}}{f_i}}}$
denote the channel gain of the diffuse link of the $i$th subcarrier \cite{Schulze_TC_2016},
where ${\eta _{\rm{D}}} = \frac{{{A_{\rm{r}}}}}{{{A_{{\rm{room}}}}}}\frac{\rho }{{1 - \rho }}$
 is the diffuse channel gain factor.
  Here, ${{A_{{\rm{room}}}}}$ is the surface of the room and
  $\rho $ is the average value of the room reflectivity factor.
Moreover, ${\tau _{\rm{D}}} =  - \frac{1}{{\ln \rho }}\frac{{4{V_{{\rm{room}}}}}}{{{A_{{\rm{room}}}}c}}$
 is the time constant,
  and ${{V_{{\rm{room}}}}}$ is the volumn of the room.
  Then, the total channel gain ${H_i}$ of the $i$th subcarrier can be expressed as
\begin{align} \label{channel_model}
{H_i} = {H_{{\rm{L}},i}} + {H_{{\rm{D}},i}},\quad i = 0,...,2N-1.
\end{align}

At the receiver, the received optical signal is converted to analog electrical signal by the PD.
Then, the digital signal is obtained by an analog-to-digital convertor (ADC).
Finally, the bit stream is recovered through the fast Fourier transform (FFT) and demodulation operations.

Specifically,
the frequency domain expression of the received signal ${Y_i}$ can be written as
\begin{align} \label{signal_model}
{Y_i} = {H_i}\sqrt {{p_i}} {X_i} + {Z_i}, \quad i = 1,...,N-1,
\end{align}
where $Y_i$ and $H_i$ represent the received signal and the total channel gain at the $i$th subcarrier, respectively.
${Z_i}$ is the additive white Gaussian noise (AWGN)  with  zero-mean, i.e., ${Z_i} \sim \mathcal{CN}\left( {0,{\sigma ^2}{W}} \right)$,
where ${\sigma ^2}$ represents the noise  power spectral density (PSD),
and the bandwidth of each subcarrier is $W$.



\begin{figure*}[!b]
    \hrulefill
    \normalsize
    \setcounter{mytempeqncnt}{\value{equation}}
    \setcounter{equation}{13}
    \begin{align}\label{total_bound_ACO}
        {R_{{\rm{L}},i}}\left( {{p_i}} \right)=W\left( {{{\log }_2}M + 1 - \frac{1}{{\ln 2}}} \right)-\sum\limits_{n = 1}^M {\frac{{W}}{M}{{\log }_2}\sum\limits_{k = 1}^M {\exp \left( { - \frac{{{p_i}{{\left| {{H_i}} \right|}^2}{{\left| {{X_{i,n}} - {X_{i,k}}} \right|}^2}}}{{2{\sigma ^2}W}}} \right)} }.
    \end{align}
\end{figure*}
\setcounter{equation}{\value{mytempeqncnt}}

\section{Achievable Rate of {the} DCO-OFDM System}
Recall that in most existing works, the achievable rates of {the} DCO-OFDM system are derived based on the assumption that
the time domain signal obtained after the IFFT is approximately Gaussian distributed,
{and the exact achievable rate without the information loss of {the} DCO-OFDM system with finite-alphabet inputs is still unknown.}

Here, we consider a system that the discrete constellation points are equiprobably drawn from  discrete
constellations set $\left\{ {{X_{i,k}}} \right\}_{k = 1}^M$
 with   cardinality $M$, where  $X_{i,k}$ is the $k$th constellation point of the $i$th subcarrier.
 Thus, the achievable rate ${R_{{\rm{F}},i}}\left( {{p_i}} \right)$ of the $i$th subcarrier can be expressed as 
\begin{subequations}\label{mutual_inf_ACO}
\begin{align}
  &{R_{{\rm{F}},i}}\left( {{p_i}} \right)  = {{I}_i}\left({X_i}; {Y_i}  \right) \label{Mutual}\\
   &= W\left( {{{\log }_2}M - \frac{1}{{\ln 2}}} \right) - \sum\limits_{n = 1}^M {\frac{{W}}{M}{\mathbb{E}_{Z_i}}\left\{ {{{\log }_2}\sum\limits_{k = 1}^M {\exp \left( { - {d_{n,k}}} \right)} } \right\}},  \label{mutual_info_new}
 \end{align}
 \end{subequations}
 where
${d_{n,k}} \buildrel \Delta \over = \frac{{{{\left| {{H_i}\sqrt {{p_i}} \left( {{X_{i,n}} - {X_{i,k}}} \right) + {Z_i}} \right|}^2}}}{{{\sigma ^2}W}}$
is a measure of the difference between discrete constellation points
$X_{i,n}$ and $X_{i,k}$,
and ${\mathbb{E}_{Z_i}}\left\{  \cdot  \right\}$ is the expectation of the noise ${Z_i}$.
The  detailed derivation   of \eqref{mutual_info_new} is given in Appendix A. Moreover,  it is easy to find that ${R_{{\rm{F}},i}}\left( {{p_i}} \right)$ is
a concave function with respect to the power allocation ${p _i}$ \cite{Xiao_TSP_2011,Rajashekar_JSAC_2019}.
Thus, the   total achievable rate of the DCO-OFDM system   with finite-alphabet inputs  is given by
\begin{align}
{R_{{\rm{F,total}}}}\left( {\left\{ {{p_i}} \right\}} \right) = \sum\limits_{i = 1}^{N - 1} {{R_{{\rm{F}},i}}\left( {{p_i}} \right)}. \label{finite_rate_total}
\end{align}

 Note that, for $M \ge 2$, the expectation term in the achievable rate \eqref{mutual_info_new}
is a non-integrable function and lacks of closed-form expression,
making it challenging for resource allocation.
To address this challenge,
 we further derive  the  lower bound of \eqref{mutual_info_new} with closed-form expression.
{Since} ${\log _2}\left( \cdot \right)$  is a concave function,
 the upper bound of the expectation term in \eqref{mutual_info_new} is given by
 \begin{subequations}
\begin{align}
&{{\mathbb{E}}_{Z_i}}\left\{ {\log _2}\sum\limits_{k = 1}^M { \exp \left( { - {d_{n,k}}} \right)}\right\}
\le {\log _2}\sum\limits_{k = 1}^M { {{\mathbb{E}}_{Z_i}} \left\{ \exp \left( { - {d_{n,k}}} \right) \right\} } \label{jense}\\
& = {\rm{lo}}{{\rm{g}}_{\rm{2}}}\sum\limits_{k = 1}^M {\int_{{Z_i}} {\exp \left( { - {d_{n,k}}} \right)\frac{1}{{\pi {\sigma ^2}W}}\exp \left( { - \frac{{{{\left| {{Z_i}} \right|}^2}}}{{{\sigma ^2}W}}} \right)} d{Z_i}}  \label{exp_z}\\
& =  - 1 + {\rm{lo}}{{\rm{g}}_{\rm{2}}}\sum\limits_{k = 1}^M {\exp \left( { - \frac{{{p_i}{{\left| {{H_i}} \right|}^2}{{\left| {{X_{i,k}} - {X_{i,m}}} \right|}^2}}}{{2{\sigma ^2}W}}} \right)}  \label{exp_upper},
\end{align}
\end{subequations}
where  \eqref{jense}  is  due to the Jensen's inequality \cite{Zeng_WCL_2012_Low}.

Thus, let ${R_{{\rm{L}},i}}\left( {{p_i}} \right)$ represent the lower bound of mutual information
of the $i$th subcarrier with finite-alphabet inputs and is given by \eqref{total_bound_ACO}.
\setcounter{equation}{14}

Similar to ${R_{{\rm{F}},i}}\left( {{p_i}} \right)$, ${R_{{\rm{L}},i}}\left( {{p_i}} \right)$ is also
a concave function with respect to the power allocation ${p _i}$.
The corresponding lower bound of total achievable rate of the DCO-OFDM system is given by
\begin{align}
{R_{{\rm{L,total}}}}\left( {\left\{ {{p_i}} \right\}} \right) = \sum\limits_{i = 1}^{N - 1} {{R_{{\rm{L}},i}}\left( {{p_i}} \right)}. \label{lower_rate_total}
\end{align}

{
As same as [36], there is a constant gap $C\triangleq\frac{1}{\ln2}-1$ between $R_{\mathrm{F}}\left(p_i\right)$ and $R_{\mathrm{L}}\left(p_i\right)$ when $p_i\to\infty$ or $p_i\to 0$, and  $R_{\mathrm{A}}\left(p_i\right)\triangleq R_{\mathrm{L}}\left(p_i\right)+C$ can also be as a low-complexity approximation of $R_{\mathrm{F}}\left(p_i\right)$, which always converges to $R_{\mathrm{F}}\left(p_i\right)$ with $p_i\to\infty$ or $p_i\to 0$, while exceeds $R_{\mathrm{F}}\left(p_i\right)$ in the medium region. Then, the corresponding total achievable rate is given by ${R_{{\rm{A,total}}}}\left( {\left\{ {{p_i}} \right\}} \right) = \sum\limits_{i = 1}^{N - 1} {{R_{{\rm{A}},i}}\left( {{p_i}} \right)}$, which is still concave with respect to the power allocation $p_i$.
}

\section{Spectral Efficiency maximization of {the} DCO-OFDM system}
In this section, we investigate the SE maximization problem of  the DCO-OFDM system
subject to the  maximum average optical power and
the maximum total electrical {transmitted} power constraints.
Specifically, we start from a simple case of ${{R_{{\rm{L,total}}}}\left( {\left\{ {{p_i}} \right\}} \right)}$  in \eqref{lower_rate_total}
with the closed-form expression %
and then the more challenging case of ${R_{{\rm{F,total}}}}\left( {\left\{ {{p_i}} \right\}} \right)$ in \eqref{finite_rate_total} without a closed-form expression.

 \subsection{SE Maximization based on ${{R_{{\rm{L,total}}}}\left( {\left\{ {{p_i}} \right\}} \right)}$}
With the  closed-form expression of  lower bound of the achievable rate    \eqref{lower_rate_total}, the SE can be expressed as 
\begin{align}  \label{lower_sub}
{\rm{S}}{{\rm{E}}_{\rm{L}}}\left( {\left\{ {{p_i}} \right\}} \right) = \frac{{{R_{{\rm{L,total}}}}\left( {\left\{ {{p_i}} \right\}} \right)}}{{2NW}}.
\end{align}
 Then, we aim to maximize the  SE of the DCO-OFDM system under amplitude constraint,
 average optical power constraint and total electrical {transmitted} power constraint,
 which can be mathematically formulated as
\begin{subequations} \label{aco_form}
\begin{align}
\mathop {{\text{maximize}}}\limits_{\left\{ {{p _i}} \right\},{I_{\rm{dc}}}} &\quad {\rm{S}}{{\rm{E}}_{\rm{L}}}\left( {\left\{ {{p _i}} \right\}} \right)  \label{PA_formu_o}\\
{\rm{s.t}}.  
& \quad {x_k} + {I_{{\rm{dc}}}} \ge 0,  \label{x_k_idc}\\
& \quad \mathbb{E}\left\{ {x_{{\rm{dc}},k}} \right\} \le {P_o},\label{dimming_aco}\\
&  \quad \sum\limits_{k = 0}^{2N - 1} {\mathbb{E}\left\{ {x_{{\rm{dc}},k}^2} \right\}}  \le {P_e}, \label{electrical_aco}\\
&  \quad {p_i} \ge 0, \quad i = 1,...,N-1.\label{PA_con3}
\end{align}
\end{subequations}
Although the joint design of variables DC-bias ${I_{\rm{dc}}}$ and power allocation ${p_i}$ complicates the optimization problem,
it can be seen from  problem \eqref{aco_form}  that the objective function is only related to the power allocation ${p_i}$,
and it is monotonically increasing as ${p_i}$ increases.
Moreover, based on \eqref{x_dc} and \eqref{x_k_idc}, the total electrical {transmitted} power in \eqref{electrical_aco} can be expressed as
\begin{subequations}
\begin{align}
\sum\limits_{k = 0}^{2N - 1} {\mathbb{E}\left\{ {x_{{\rm{dc}},k}^2} \right\}}  &= \sum\limits_{k = 0}^{2N - 1} {\mathbb{E}\left\{ {x_k^2} \right\}}  + \sum\limits_{k = 0}^{2N - 1} {\mathbb{E}\left\{ {I_{{\rm{dc}}}^2} \right\}} \label{elec_cons_1} \\
 &  = 2\sum\limits_{i = 1}^{N - 1} {{p_i}}  + 2N\mathbb{E}\left\{ {I_{{\rm{dc}}}^2} \right\} ,\label{elec_cons_2}
\end{align}
\end{subequations}
where  \eqref{elec_cons_1} is true since the time domain signal ${x_k}$ is not clipped, i.e., ${x_{{\rm{clip}},k}} = {x_k}$,
\eqref{elec_cons_2} follows from the  Parseval's theorem, and the total electrical {transmitted} power of time domain signal $x_k$ is $\sum\nolimits_{k = 0}^{2N - 1} {\mathbb{E}\left\{ {x_k^2} \right\}}  = \sum\nolimits_{i = 0}^{2N - 1} {{p_i}}  = 2\sum\nolimits_{i = 1}^{N - 1} {{p_i}} $.
Thus, combining with the total electrical {transmitted} power constraint, we find that more power should be allocated for the information-carrying ${p _i}$ and less power should be allocated for the DC-bias ${I_{\rm{dc}}}$ to maximize the achievable rate ${R_{{\rm{L,total}}}}\left( {\left\{ {{p_i}} \right\}} \right)$. 
Specifically, the optimal value of $I_{{\rm{dc}}}^{{\rm{opt}}}$  should be the minimum DC-bias without clipping the signal ${x_k}$. 

Furthermore, from \eqref{IFFT_1}, we have
\begin{align}
{x_k} \ge  - \sqrt {\frac{2}{N}}\sum\limits_{i = 1}^{N - 1} {\sqrt {{p_i}} \left| {{X_i}} \right|}. \label{xk_min}
\end{align}
Then, { $I_{{\rm{dc}}}^{{\rm{non-clipping}}}$} which minimizes the power of the DC-bias without clipping the signal ${x_k}$ can be written as
\begin{align}
{ I_{{\rm{dc}}}^{{\rm{non-clipping}}}} = \sqrt {\frac{2}{N}}\sum\limits_{i = 1}^{N - 1} {\sqrt {{p_i}} \left| {{X_i}} \right|}. \label{Idc_optimal}
\end{align}
As a result, combining \eqref{x_dc}, \eqref{x_k_idc} and \eqref{Idc_optimal},  the  average optical power can be rewritten as
\begin{subequations}
\begin{align}
 \mathbb{E}\left\{ {x_{{\rm{dc}},k}} \right\} &= \mathbb{E}\left\{ {x_{{\rm{clip}},k}} + { I_{{\rm{dc}}}^{{\rm{non-clipping}}}}  \right\} \label{dimming_c_2} \\
 &=   \mathbb{E}\left\{ {{x_k}} \right\} + \mathbb{E}\left\{ {{ I_{{\rm{dc}}}^{{\rm{non-clipping}}}}} \right\}  \label{dimming_c_3} \\
  &=\sqrt {\frac{2}{N}} \sum\limits_{i = 1}^{N - 1} {\sqrt {{p_i}} \mathbb{E}\left\{ {\left| {{X_i}} \right|} \right\}}. \label{dimming_c_1}
 \end{align}
 \end{subequations}
According to Cauchy-Schwarz inequality
\begin{align} \label{inequality}
{\left( {\sum\limits_{i = 1}^n {{a_i}} } \right)^2} \le n\sum\limits_{i = 1}^n {a_i^2} ,\quad {a_i} \ge 0 ,
\end{align}
the upper bound of the square of the average optical power is given by
\begin{align}
{\left( {\sqrt {\frac{2}{N}} \sum\limits_{i = 1}^{N - 1} {\sqrt {{p_i}} \mathbb{E}\left\{ {\left| {{X_i}} \right|} \right\}} } \right)^2} \le \frac{{2\left( {N - 1} \right)}}{N}\sum\limits_{i = 1}^{N - 1} {{p_i}} {\mathbb{E}^2}\left\{ {\left| {{X_i}} \right|} \right\}.
\end{align}
Thus, the average  optical power constraint \eqref{dimming_aco} can be restricted {to}
\begin{align} \label{MI_con3_jian}
\sum\limits_{i = 1}^{N - 1} {{p_i}} {\mathbb{E}^2}\left\{ {\left| {{X_i}} \right|} \right\} \le \frac{{NP_o^2}}{{2\left( {N - 1} \right)}}.
\end{align}

Meanwhile, by substituting ${ I_{{\rm{dc}}}^{{\rm{non-clipping}}}}$  in \eqref{Idc_optimal} into \eqref{elec_cons_2}, the total electrical {transmitted} power can be written as
\begin{align}
\sum\limits_{k = 0}^{2N - 1} {\mathbb{E}\left\{ {x_{{\rm{dc}},k}^2} \right\}}  = 2\sum\limits_{i = 1}^{N - 1} {{p_i}}  + 4\mathbb{E}\left\{ {{{\left( {\sum\limits_{i = 1}^{N - 1} {\sqrt {{p_i}} \left| {{X_i}} \right|} } \right)}^2}} \right\}. \label{PA_con1_new}
\end{align}
Applying the  inequality \eqref{inequality}, the upper bound of the total electrical {transmitted} power \eqref{PA_con1_new} is given as
\begin{align}
\sum\limits_{k = 0}^{2N - 1} {\mathbb{E}\left\{ {x_{{\rm{dc}},k}^2} \right\}}  \le 2\sum\limits_{i = 1}^{N - 1} {{p_i}}  + 4\mathbb{E}\left\{ {\left( {N - 1} \right)\sum\limits_{i = 1}^{N - 1} {{p_i}{{\left| {{X_i}} \right|}^2}} } \right\}.
\end{align}
Thus, the constraint \eqref{electrical_aco} can be reformulated as
\begin{align} \label{constraint_idc}
\sum\limits_{i = 1}^{N - 1} {{p_i}}  \le \frac{{{P_e}}}{{4N - 2}}.
\end{align}

Therefore, the SE maximization  problem \eqref{aco_form}  can be transformed into
\begin{subequations}   \label{finite_lowerbound_ACO}
\begin{align}
\mathop {{\text{maximize}}}\limits_{\left\{ {{p _i}} \right\}} &\quad {\rm{S}}{{\rm{E}}_{\rm{L}}}\left( {\left\{ {{p _i}} \right\}} \right)   \label{aco_lower_a}\\
{\rm{s.t. }} &\quad \sum\limits_{i = 1}^{N - 1} {{p_i}} {\mathbb{E}^2}\left\{ {\left| {{X_i}} \right|} \right\} \le \frac{{NP_o^2}}{{2\left( {N - 1} \right)}}, \label{power_con1}\\
&\quad \sum\limits_{i = 1}^{N - 1} {{p_i}}  \le \frac{{{P_e}}}{{4N - 2}},  \label{power_con2}\\
&\quad  p_i \ge 0{\rm{ }},\quad i = 1,...,N-1. \label{power_con3}
\end{align}
\end{subequations}

Note that the optimization problem \eqref{finite_lowerbound_ACO} has a strictly concave objective function over its input power ${p_i}$
 and linear constraints, and
thus can be efficiently solved by the interior-point algorithm. {Such as the barrier method, it transforms the convex optimization problems into a sequence of equality constrained problems and applies Newton's method to them, or such as the primal-dual interior-point method, it modifies the corresponding KKT conditions and solves them by Newton's method\cite{Boyd_Convex_2004,joseph2006}. Besides, It has been implemented by standard convex optimization solvers such as CVX \cite{cvx}.}

{
Besides, the SE based on $R_{{\rm{L,total}}}\left( {\left\{ {{p_i}} \right\}} \right)$ is denoted as $\mathrm{SE}_{\mathrm{A}}$, and the corresponding SE  maximization power allocation problem is equivalent to the problem (17) without the optimality loss, wich can be solved by the similar technique.  
}

\subsection{SE Maximization based on ${{R_{{\rm{F,total}}}}\left( {\left\{ {{p_i}} \right\}} \right)}$}

In this subsection, we investigate the SE maximization  problem of the DCO-OFDM system   based on the exact mutual information \eqref{finite_rate_total}.
The SE achieved with exact mutual information is given by
\begin{align}
{\rm{S}}{{\rm{E}}_{\rm{F}}}\left( {\left\{ {{p_i}} \right\}} \right) = \frac{{R_{{\rm{F,total}}}}\left( {\left\{ {{p_i}} \right\}} \right)}{{2NW}}.
\end{align}

Similarly, the value of the DC-bias ${I_{{\rm{dc}}}}$ is the same as \eqref{Idc_optimal}
to ensure that the transmitted DC-bias power is minimized while the signal ${x_k}$ is not clipped.
Thus, by considering the same constraints in problem  \eqref{finite_lowerbound_ACO},
 the SE maximization problem with finite-alphabet inputs can be reformulated as
\begin{align}   \label{aco_finite}
\mathop {{\text{maximize}}}\limits_{\left\{ {{p _i}} \right\}} &\quad {\rm{S}}{{\rm{E}}_{\rm{F}}}\left( {\left\{ {{p _i}} \right\}} \right) \\
{\rm{s}}{\rm{.t}}{\rm{. }} &\quad \eqref{power_con1},\eqref{power_con2},\eqref{power_con3} \nonumber.
\end{align}

Note that the objective function \eqref{aco_finite} is concave over ${p _i}$,
and constraints \eqref{power_con1} and \eqref{power_con2} are affine functions over ${p _i}$.
This type of  optimization problem can be efficiently solved based on the KKT conditions.
 To this end, we first derive
 the  Lagrangian function of   problem \eqref{aco_finite}, which is given by
\begin{align}
{\mathcal{L}_{\rm{F}}} =  - \sum\limits_{i = 1}^{N - 1} {{R_{{\rm{F}},i}}\left( {{p_i}} \right)}  &+ {\lambda _1}\left( {\sum\limits_{i = 1}^{N - 1} {{p_i}{\mathbb{E}^2}\left\{ {\left| {{X_i}} \right|} \right\}}  - \frac{{NP_o^2}}{{2\left( {N - 1} \right)}}} \right){\rm{ }} \nonumber\\
&+ {\lambda _2}\left( {\sum\limits_{i = 1}^{N - 1} {{p_i}}  - \frac{{{P_e}}}{{4N - 2}}} \right),
\end{align}
where ${\lambda _1} \ge 0$, ${\lambda _2} \ge 0$
are the Lagrange multipliers corresponding to constraint \eqref{power_con1} and  \eqref{power_con2} respectively.
Then, the KKT conditions of problem \eqref{aco_finite}  are given  as
\begin{subequations}
\begin{align}
&\quad \frac{{\partial {\mathcal{L}_{\rm{F}}}}}{{\partial {p_i}}} =  - \frac{{\partial {R_{{\rm{F}},i}}\left( {{p_i}} \right)}}{{\partial {p_i}}} + {\lambda _1}{\mathbb{E}^2}\left\{ {\left| {{X_i}} \right|} \right\} + {\lambda _2} = 0,\label{acoKKT_1}\\
&\quad {\lambda _1}\left( {\sum\limits_{i = 1}^{N - 1} {{p_i}{\mathbb{E}^2}\left\{ {\left| {{X_i}} \right|} \right\}}  - \frac{{NP_o^2}}{{2\left( {N - 1} \right)}}} \right) = 0,\label{acoKKT_2}\\
&\quad {\lambda _2}\left( {\sum\limits_{i = 1}^{N - 1} {{p_i}}  - \frac{{{P_e}}}{{4N - 2}}} \right) = 0,\label{acoKKT_3}\\
&\quad \sum\limits_{i = 1}^{N - 1} {{p_i}{\mathbb{E}^2}\left\{ {\left| {{X_i}} \right|} \right\}}  - {\frac{{NP_o^2}}{{2\left( {N - 1} \right)}}} \le 0,\\
&\quad \sum\limits_{i = 1}^{N - 1} {{p_i}}  - \frac{{{P_e}}}{{4N - 2}} \le 0,\\
&\quad {\lambda _1} \ge 0,\quad {\lambda _2} \ge 0, \quad p_i \ge 0, \quad i = 1,...,N-1.
\end{align}
\end{subequations}

However, it is challenging to directly calculate the partial derivative in \eqref{acoKKT_1}
due to the lack of the closed-form expressions for the achievable rate ${R_{{\rm{F}},i}}\left( {{p_i}} \right)$.
Therefore, we aim to address this difficulty by exploiting the relationship between the mutual information and MMSE.
Specifically, the  MMSE  of  ${X_i}$ is given as
\begin{align}\label{MMSE1}
{\rm{MMS}}{{\rm{E}}_i}\left( {{\rm{SN}}{{\rm{R}}_i}} \right) = \mathbb{E}\left\{ {{{\left| {{X_i} - {{\hat X}_i}} \right|}^2}} \right\},
\end{align}
where ${\rm{SN}}{{\rm{R}}_i} = \frac{{{{\left| {{H_i}} \right|}^2}{p_i}}}{{{\sigma ^2}W}}$ is the signal-to-noise ratio of the $i$th subcarrier,
and ${{\hat X}_i}$ is conditional expectation of  ${X_i}$, i.e., ${{\hat X}_i} = \mathbb{E}\left\{ {{X_i}\left| {{Y_i} = {H_i}\sqrt {{p _i}}  {X_i} + {Z_i}} \right.} \right\}$.
According to  \emph{Theorem 1} in \cite{Guo_TIT_2005},
 the  relationship between the  mutual information \eqref{Mutual}  and the  MMSE \eqref{MMSE1} is given by
\begin{align} \label{MMSE_MI_aco}
\frac{\partial }{{\partial {\rm{SN}}{{\rm{R}}_i}}}{I_i}\left( {{X_i};{Y_i}} \right) = {\rm{MMS}}{{\rm{E}}_i}\left( {{\rm{SN}}{{\rm{R}}_i}} \right).
\end{align}
Combining \eqref{mutual_info_new} and \eqref{MMSE_MI_aco},
partial derivative of function ${R_{{\rm{F}},i}}\left( {{p_i}} \right)$ can be written as
\begin{align}\label{MMSE_MI_aco_2}
\frac{{\partial {R_{{\rm{F}},i}}\left( {{p_i}} \right)}}{{\partial {p_i}}} = \frac{{{{\left| {{H_i}} \right|}^2}}}{{{\sigma ^2}W}}{\rm{MMS}}{{\rm{E}}_i}\left( {\frac{{{{\left| {{H_i}} \right|}^2}}}{{{\sigma ^2}W}}{p_i}} \right).
\end{align}

By substituting \eqref{MMSE_MI_aco_2} into \eqref{acoKKT_1},
 we have
 \begin{align} \label{MMSE_p}
\frac{{{{\left| {{H_i}} \right|}^2}}}{{{\sigma ^2}W}}{\rm{MMS}}{{\rm{E}}_i}\left( {\frac{{{{\left| {{H_i}} \right|}^2}}}{{{\sigma ^2}W}}{p_i}} \right) = {\lambda _1}{\mathbb{E}^2}\left\{ {\left| {{X_i}} \right|} \right\} + {\lambda _2}.
 \end{align}

Then, according to \eqref{MMSE_p}, the  power allocation $p _i$ can be obtained as
\begin{align}   \label{alpha_aco}
{p_i} = \frac{{{\sigma ^2}W}}{{{{\left| {{H_i}} \right|}^2}}}{\rm{MMSE}}_i^{ - 1}\left[ {\frac{{{\sigma ^2}W}}{{{{\left| {{H_i}} \right|}^2}}}\left( {{\lambda _1}{\mathbb{E}^2}\left\{ {\left| {{X_i}} \right|} \right\} + {\lambda _2}} \right)} \right],
\end{align}
where ${\rm{MMSE}}_i^{ - 1}\left(  \cdot  \right)$ is
  the inverse function of ${\rm{MMS}}{{\rm{E}}_i}\left(  \cdot  \right)$
with domain in $\left[ {0,1} \right]$ and ${\rm{MMSE}}_i^{ - 1}\left( 1 \right) = 0$ \cite{Lozano_TIT_2006}.
 Therefore, for the  SE maximization problem    \eqref{aco_finite}, the optimal allocation  power  of the $i$th subcarrier    is given by \eqref{ai}.
\setcounter{equation}{38}


\begin{figure*}[!b]
    \hrulefill
    \normalsize
    \setcounter{mytempeqncnt}{\value{equation}}
    \setcounter{equation}{37}
    \begin{align} \label{ai}
        p_i^{{\rm{opt}}} = \left\{ {\begin{array}{*{20}{c}}
                {\frac{{{\sigma ^2}W}}{{{{\left| {{H_i}} \right|}^2}}}{\rm{MMSE}}_i^{ - 1}\left[ {\frac{{{\sigma ^2}W}}{{{{\left| {{H_i}} \right|}^2}}}\left( {{\lambda _1}{\mathbb{E}^2}\left\{ {\left| {{X_i}} \right|} \right\} + {\lambda _2}} \right)} \right],}&{0 < {\lambda _1}{\mathbb{E}^2}\left\{ {\left| {{X_i}} \right|} \right\} + {\lambda _2} \le \frac{{{{\left| {{H_i}} \right|}^2}}}{{{\sigma ^2}W}},}\\
                {0,}&{{\rm{otherwise}}.}
        \end{array}} \right.
    \end{align}

\setcounter{equation}{39}
\begin{align}\label{auxi_G}
    G_i\left(\lambda_1, \lambda_2\right)\triangleq
    \begin{cases}
        \frac{\left|H_i\right|^2}{\sigma^2W\left(\lambda_1\mathbb{E}^2\left\{\left|X_i\right|\right\}+\lambda_2\right)}
        -\mathrm{MMSE}^{-1}\left[\frac{\sigma^2W}{\left|H_i\right|^2}\left(\lambda_1\mathbb{E}^2\left\{\left|X_i\right|\right\}+\lambda_2\right)\right],&0<\lambda_1\mathbb{E}^2\left\{\left|X_i\right|\right\}+\lambda_2\leq \frac{\sigma^2W}{\left|H_i\right|^2};\\
        1, & \mathrm{otherwise}.
    \end{cases}
\end{align}

\end{figure*}
\setcounter{equation}{\value{mytempeqncnt}}

{
    Meanwhile, according to the definition of MMSE, $\lambda_1$ and $\lambda_2$ should be greater than $0$, otherwise the required power would be infinity. Therefore, substituting \eqref{ai} into the complementary slackness conditions \eqref{acoKKT_2} and \eqref{acoKKT_3}, the dual variables $\lambda_1$ and $\lambda_2$ is the solution of the following equation:
    \begin{subequations}
        \begin{align}
            &\sum\limits_{i = 1}^{N - 1} {{{\frac{{{\sigma ^2}W}}{{{{\left| {{H_i}} \right|}^2}}}{\rm{MMSE}}_i^{ - 1}\left[ {\frac{{{\sigma ^2}W}}{{{{\left| {{H_i}} \right|}^2}}}\left( {{\lambda _1}{\mathbb{E}^2}\left\{ {\left| {{X_i}} \right|} \right\} + {\lambda _2}} \right)} \right]}}{\mathbb{E}^2}\left\{ {\left| {{X_i}} \right|} \right\}}  \nonumber\\
                &\qquad\qquad\qquad\qquad\qquad\qquad\qquad\qquad= \frac{{NP_o^2}}{{2\left( {N - 1} \right)}},\\
            &{\sum\limits_{i = 1}^{N - 1} {{{\frac{{{\sigma ^2}W}}{{{{\left| {{H_i}} \right|}^2}}}{\rm{MMSE}}_i^{ - 1}\left[ {\frac{{{\sigma ^2}W}}{{{{\left| {{H_i}} \right|}^2}}}\left( {{\lambda _1}{\mathbb{E}^2}\left\{ {\left| {{X_i}} \right|} \right\} + {\lambda _2}} \right)} \right]}}}  = \frac{{{P_e}}}{{4N - 2}}},
        \end{align}
    \end{subequations}
    which can obtained by the multi-level mercury-water-filling power allocation scheme as listed in Algorithm 1 \cite{Zhang_2008_JSTSP, Lozano_TIT_2006}.
}
{
    Moreover, to facilitate the explanation of the power allocation scheme, an auxiliary function $G_i\left(\lambda_1, \lambda_2\right)$ is defined as follows \eqref{auxi_G}.
\setcounter{equation}{40}Then, the allocated power $p_i$ can be represented as
    \begin{align}\label{mercury-water-filling-m}
        p_i=\frac{1}{\lambda_1\mathbb{E}^2\left\{\left|X_i\right|\right\}+\lambda_2}-\frac{\left|H_i\right|^2}{\sigma^2W}G_i\left(\lambda_1, \lambda_2\right).
    \end{align} 
}


\begin{algorithm}[htbp]
    \caption{Multi-level Mercury-water-filling  Power Allocation Scheme}
    \label{Mutilevel_Bisection Method_alg}
    \begin{algorithmic}[1]
        \Require  Given ${\lambda _1} \in \left[ {0,{{\hat \lambda }_1}} \right]$, ${\delta _1} > 0$, where ${{{\hat \lambda }_1}}$ is the upper bound of ${\lambda_1}$ and ${\delta _1}$ is a small positive constant that controls the algorithm accuracy. Initialize  ${\lambda _{\min }} = 0,{\lambda _{\max }} = {{\hat \lambda }_1}$;
        \While{${\lambda _{\max }} - {\lambda _{\min }} \ge {\delta _1}$}
            \State Set ${\lambda _1} = {{\left( {{\lambda _{\min }} + {\lambda _{\max }}} \right)} \mathord{\left/
 {\vphantom {{\left( {{\lambda _{\min }} + {\lambda _{\max }}} \right)} 2}} \right.
 \kern-\nulldelimiterspace} 2}$;
            \State Find the minimum ${\lambda _2} \ge 0$, with which
            \Statex \quad \quad \quad $\sum\limits_{i = 1}^{N - 1} {{{\left[ {\frac{{{\sigma ^2}W}}{{{{\left| {{H_i}} \right|}^2}}}{\rm{MMSE}}_i^{ - 1}\left[ {\frac{{{\sigma ^2}W}}{{{{\left| {{H_i}} \right|}^2}}}\left( {{\lambda _1}{\mathbb{E}^2}\left\{ {\left| {{X_i}} \right|} \right\} + {\lambda _2}} \right)} \right]} \right]}^ + }}  \le \frac{{{P_e}}}{{4N - 2}}$;
          \State If ${\lambda _1}{\mathbb{E}^2}\left\{ {\left| {{X_i}} \right|} \right\} + {\lambda _2} \le {\left| {{H_i}} \right|^2}/{\sigma ^2}W$,
                  substitute ${\lambda _1},{\lambda _2}$ to  obtain
           \Statex \quad \quad \quad $p_i^{{\rm{opt}}} = \frac{{{\sigma ^2}W}}{{{{\left| {{H_i}} \right|}^2}}}{\rm{MMSE}}_i^{ - 1}\left[ {\frac{{{\sigma ^2}W}}{{{{\left| {{H_i}} \right|}^2}}}\left( {{\lambda _1}{\mathbb{E}^2}\left\{ {\left| {{X_i}} \right|} \right\} + {\lambda _2}} \right)} \right]$;
             otherwise $p_i^{{\rm{opt}}} = 0$;
          \State If $\sum\limits_{i = 1}^{N - 1} {{p_i^{{\rm{opt}}}}{\mathbb{E}^2}\left\{ {\left| {{X_i}} \right|} \right\}}  \le {\frac{{NP_o^2}}{{2\left( {N - 1} \right)}}}$,
             set ${\lambda _{\max }} \leftarrow {\lambda _1}$; otherwise ${\lambda _{\min }} \leftarrow {\lambda _1}$;
        \EndWhile
        \Ensure $p_i^{{\rm{opt}}}$;
    \end{algorithmic}
\end{algorithm}

\section{Energy Efficiency maximization of {the} DCO-OFDM system}

In this section,
we propose the optimal power allocation schemes to maximize   the EE of  the DCO-OFDM  system subject to  the minimum SE  threshold,
the maximum average optical {power and} the maximum total electrical {transmitted} power constraints.
We  first investigate the EE maximization problem  based on the lower bound ${{R_{{\rm{L,total}}}}\left( {\left\{ {{p_i}} \right\}} \right)}$
in \eqref{lower_rate_total} with the closed-form expression, and then study the  case based on the exact achievable rate
  ${R_{{\rm{F,total}}}}\left( {\left\{ {{p_i}} \right\}} \right)$ in \eqref{finite_rate_total} without a closed-form expression.

\subsection{EE Maximization based on ${{R_{{\rm{L,total}}}}\left( {\left\{ {{p_i}} \right\}} \right)}$}
Based on the rate expression \eqref{lower_rate_total}, the EE is  given by
\begin{align}     \label{EE_ACO}
{\rm{E}}{{\rm{E}}_{\rm{L}}}\left( {\left\{ {{p_i}} \right\}} \right) = \frac{{{R_{{\rm{L}},{\rm{total}}}}\left( {\left\{ {{p_i}} \right\}} \right)}}{{2\sum\limits_{i = 1}^{N - 1} {{p_i}}  + {P_{{\rm{dc}}}} + {P_c}}},
\end{align}
where ${P_{{\rm{dc}}}} = \sum\nolimits_{k = 0}^{2N - 1} {\mathbb{E}\left\{ {{{\left( {{I_{{\rm{dc}}}}} \right)}^2}} \right\}}$
 denotes the DC-bias power
and ${P_c}$ is the constant total circuit consumption of the whole system.

Therefore,  the  corresponding EE maximization problem can be formulated  as
\begin{subequations}  \label{EE_ACO_lower1}
\begin{align}
\mathop {{\rm{maximize}}}\limits_{\left\{ {{p_i}} \right\},{I_{{\rm{dc}}}}} &\quad{\rm{E}}{{\rm{E}}_{\rm{L}}}\left( {\left\{ {{p_i}} \right\}} \right)   \label{EE_lowerbound_ACO}\\
 {\rm{s}}{\rm{.t}}{\rm{. }}   &\quad \eqref{x_k_idc},\eqref{dimming_aco},\eqref{electrical_aco}, \eqref{PA_con3},   \nonumber \\
 &\quad {\rm{S}}{{\rm{E}}_{\rm{L}}}\left( {\left\{ {{p_i}} \right\}} \right) \ge \gamma,  \label{EE_lowerbound_ACO_con12}
\end{align}
\end{subequations}
where $\gamma  = \frac{{\bar R}}{{2NW}}$ is the minimum SE requirement  of the DCO-OFDM system,
and ${\bar R}$ is the corresponding minimum threshold of the total achievable rate.

To ensure that there is no loss of information during the clipping operation,
 the   value of    ${I_{{\rm{dc}}}}$ is given in \eqref{Idc_optimal}, and ${P_{{\rm{dc}}}}$ is given as
 \begin{align}\label{P_dc_org}
{P_{{\rm{dc}}}} = \sum\limits_{k = 0}^{2N - 1} {\mathbb{E}\left\{ {{{\left( {{ I_{{\rm{dc}}}^{{\rm{non-clipping}}}}} \right)}^2}} \right\}}  = 4\mathbb{E}\left\{ {{{\left( {\sum\limits_{i = 1}^{N - 1} {\sqrt {{p_i}} \left| {{X_i}} \right|} } \right)}^2}} \right\}.
\end{align}

Thus, the original joint optimization problem \eqref{EE_ACO_lower1}
is converted into a concave-concave fractional problem, which is still complex and hard to solve.
To overcome this challenge, ${P_{{\rm{dc}}}}$ in \eqref{P_dc_org} is reformulated
based on the Cauchy-Schwarz inequality \eqref{inequality} and is given by
\begin{align}\label{P_dc}
{P_{{\rm{dc}}}} \le 4\mathbb{E}\left\{ {\left( {N - 1} \right)\sum\limits_{i = 1}^{N - 1} {{p_i}{{\left| {{X_i}} \right|}^2}} } \right\} = 4\left( {N - 1} \right)\sum\limits_{i = 1}^{N - 1} {{p_i}}.
\end{align}
Based on \eqref{P_dc}, the denominator of \eqref{EE_ACO} can be treated as an affine function of ${p_i}$.
Thus, problem \eqref{EE_ACO_lower1} can be transformed into a concave-linear fractional problem with variable ${p_i}$.

Similar to the SE maximization problem,  the  average optical power constraint \eqref{dimming_aco}
and total electrical {transmitted} power constraint \eqref{electrical_aco} can be reformulated as \eqref{power_con1} and \eqref{power_con2}, respectively.
Therefore, the   EE maximization problem  \eqref{EE_ACO_lower1} can be reformulated  as
\begin{subequations}  \label{EE_ACO_lower}
\begin{align}
\mathop {{\rm{maximize}}}\limits_{\left\{ {{p_i}} \right\}} &\quad \frac{{{R_{{\rm{L}},{\rm{total}}}}\left( {\left\{ {{p_i}} \right\}} \right)}}{{\left( {4N - 2} \right)\sum\limits_{i = 1}^{N - 1} {{p_i}}  + {P_c}}}   \label{EE_lowerbound_ACO_2}\\
 {\rm{s.t.}}  &\quad  \eqref{power_con1},\eqref{power_con2},\eqref{power_con3},   \nonumber \\
 &\quad {\rm{S}}{{\rm{E}}_{\rm{L}}}\left( {\left\{ {{p_i}} \right\}} \right) \ge \gamma.  \label{EE_lowerbound_ACO_con2}
\end{align}
\end{subequations}

Note that those constraints of problem \eqref{EE_ACO_lower} form a convex feasible solution set.
It can be seen from the objective function \eqref{EE_lowerbound_ACO_2} that the numerator ${R_{{\rm{L,total}}}}\left( {\left\{ {{p_i}} \right\}} \right)$
is  concave  over its input power
and  the denominator is an affine function of ${p_i}$.
In the following, we employ Dinkelbach-type iterative algorithm \cite{Dinkelbach,Crouzeix_1991} to handle this
concave-linear fractional problem
by converting problem \eqref{EE_ACO_lower} into a sequence of convex sub-problems.
In particular, by iteratively solving these convex subproblems,
the globally optimal solution of problem \eqref{EE_ACO_lower} can be obtained eventually \cite{Zappone_2015}.

Let us define  a new function $f\left( {\left\{ {{p_i}} \right\},q} \right)$ as follow
 \begin{align} \label{Dinkelbach_f}
f\left( {\left\{ {{p_i}} \right\},q} \right){\rm{ }} \buildrel \Delta \over = {R_{{\rm{L}},{\rm{total}}}}\left( {\left\{ {{p_i}} \right\}} \right) - q\left[ {\left( {4N - 2} \right)\sum\limits_{i = 1}^{N - 1} {{p_i}}  + {P_c}} \right],
 \end{align}
where $q$ is a real parameter to be found iteratively.
When $q$ is as large as possible, the optimal solution of problem \eqref{EE_ACO_lower} can be obtained
 by calculating the roots of the equation
$f\left( {\left\{ {{p_i}} \right\},q} \right) = 0$ in the feasible constraint set \cite{Zappone_2015}.

For a given $q$ in each iteration, the convex subproblem over ${{p_i}}$ can be expressed as
\begin{align}    \label{dinkelbach_DCO}
  \mathop {{\text{maximize}}
 }\limits_{\left\{ {{p _i}} \right\}} &\quad f\left( {\left\{ {{p_i}} \right\},q} \right)\\
  {\rm{s}}{\rm{.t}}{\rm{. }}   &\quad \eqref{power_con1},\eqref{power_con2},\eqref{power_con3},\eqref{EE_lowerbound_ACO_con2}.   \nonumber
 \end{align}

Since the optimization problem \eqref{dinkelbach_DCO} has a concave objective function over its input power and linear constrains,
 the optimal power $p_i^{{\rm{opt}}}$ of problem \eqref{dinkelbach_DCO} can be obtained by the interior-point algorithm. {Such as the barrier method, it transforms the convex optimization problems into a sequence of equality constrained problems and applies Newton's method to them, or such as the primal-dual interior-point method, it modifies the corresponding KKT conditions and solves them by Newton's method\cite{Boyd_Convex_2004,joseph2006}. Besides, It has been implemented by standard convex optimization solvers such as CVX \cite{cvx}.}

Finally, the EE maximization problem for the lower bound of achievable rate
 can be solved by the Dinkelbach-type algorithm.
The Dinkelbach-type algorithm is guaranteed to converge to the optimal solution of problem \eqref{EE_ACO_lower}
 with a finite number of iterations \cite{Dinkelbach,Zappone_2015,Crouzeix_1991}.
The details of  implementation are shown in  algorithm \ref{dinkelbach_alg}.

 {
Similarly, the $\mathrm{EE}_{\mathrm{A}}$ can be defined with $R_{{\rm{L,total}}}\left( {\left\{ {{p_i}} \right\}} \right)$, and the corresponding EE  maximization power allocation problem can be built and solved as similar to the problem (43).
}

\begin{algorithm}[htbp]
    \caption{Dinkelbach-type  Power Allocation Scheme}
    \label{dinkelbach_alg}
    \begin{algorithmic}[1]
        \Require  Given $\delta  \to 0,n = 0, p_i^{{\rm{opt}}} > 0,{q ^{\left( n \right)}} = 0$;
        \While{$\left| {{q ^{\left( n \right)}} - {q ^{\left( {n + 1} \right)}}} \right| \le \delta $}
            \State Compute the optimal solution $p_i^{{\rm{opt}}}$ in \eqref{dinkelbach_DCO};
            \State Calculating the value of function $f\left( {\left\{ {p_i^{{\rm{opt}}}} \right\},{q^{\left( n \right)}}} \right)$;
            \State ${q^{\left( {n + 1} \right)}} = {\rm{E}}{{\rm{E}}_{\rm{L}}}\left( {\left\{ {p_i^{{\rm{opt}}}} \right\}} \right)$;
            \State $n = n + 1$;
        \EndWhile
        \Ensure ${\rm{E}}{{\rm{E}}_{\rm{L}}}\left( {\left\{ {p_i^{{\rm{opt}}}} \right\}} \right)$.
    \end{algorithmic}
\end{algorithm}

\subsection{EE Maximization based on ${{R_{{\rm{F,total}}}}\left( {\left\{ {{p_i}} \right\}} \right)}$}
By applying the exact achievable rate  expression given by \eqref{finite_rate_total},
 the EE of finite-alphabet inputs can be expressed as
\begin{align}     \label{EE_F_long}
{\rm{E}}{{\rm{E}}_{\rm{F}}}\left( {\left\{ {{p_i}} \right\}} \right) = \frac{{{R_{{\rm{F}},{\rm{total}}}}\left( {\left\{ {{p_i}} \right\}} \right)}}{{2\sum\limits_{i = 1}^{N - 1} {{p_i}}  + {P_{{\rm{dc}}}} + {P_c}}}.
\end{align}
Then,  the EE maximization  problem with finite-alphabet inputs can be reformulated as
 \begin{subequations}    \label{finite_EE_ACO}
 \begin{align}
  \mathop {{\text{maximize}}
 }\limits_{\left\{ {{p _i}} \right\}} &\quad {\rm{E}}{{\rm{E}}_{\rm{F}}}\left( {\left\{ {{p _i}} \right\}} \right)  \label{finite_EE_ACO_a}\\
  {\rm{s}}{\rm{.t}}{\rm{. }}   &\quad \eqref{power_con1},\eqref{power_con2},\eqref{power_con3},   \nonumber \\
  &\quad {\rm{S}}{{\rm{E}}_{\rm{F}}}\left( {\left\{ {{p_i}} \right\}} \right) \ge \gamma . \label{finite_EE_ACO_con1}
 \end{align}
 \end{subequations}

Here, ${P_{{\rm{dc}}}}$ in the denominator of \eqref{finite_EE_ACO_a} takes the same value and constraint as \eqref{P_dc_org} and \eqref{P_dc}.
Due to the concave numerator ${R_{{\rm{F,total}}}}\left( {\left\{ {{p_i}} \right\}} \right)$ in \eqref{finite_EE_ACO_a},
problem \eqref{finite_EE_ACO} is  also a concave-linear fractional problem,
which can also be solved by Dinkelbach-type algorithms.

\section{Simulation Results and Discussion}

In this section, we present numerical results to illustrate the proposed  power allocation schemes  for the SE  and EE maximization problems of the DCO-OFDM VLC system.
We consider the above DCO-OFDM VLC system is in a $\left( {5 \times 5 \times 3} \right){{\rm{m}}^3}$ room equipped with four LED lights,
and the origin  $\left( {0, 0, 0} \right)$ of the three-dimensional Cartesian coordinate system $\left( {X, Y, Z} \right)$
is located at one corner on the floor of the square room.
The receiver is located at $\left( {0.5, 1, 0} \right)$m,
and the  four LEDs are respectively located at
$\left( {1.5, 1.5, 3} \right)$m, $\left( {1.5, 3.5, 3} \right)$m, $\left( {3.5, 1.5, 3} \right)$m, and $\left( {3.5, 3.5, 3} \right)$m.
The other basic parameters of the system are listed in Table \ref{baisc_par}. 

\begin{table}[htbp]
    \centering
    \caption{Simulation Parameters of  the  DCO-OFDM VLC System.}\label{baisc_par}
    \begin{tabular}{|l|l|}
        \hline
        \textbf{Definition}  & \textbf{Value}   \\
        \hline
        Half the number of subcarriers, $N$ & $16$   \\
        \hline
        Room size & $\left( {5 \times 5 \times 3} \right){{\rm{m}}^3}$ \\
        \hline
        FOV, $\Psi$ & ${90^ \circ }$   \\
        \hline
        Lambertian emission order, $m$ & 1   \\
        \hline
        Half power angle, ${\Phi _{1/2}}$ & ${60^ \circ }$ \\
        \hline
        PD collection area, ${A_{\rm{r}}}$ & $1~{\rm{c}}{{\rm{m}}^2}$  \\
        \hline
        Reflectivity factor, $\rho $ & 0.8 \\
        \hline
        Circuit power consumption,  $P_c$   &   $0.1~{\rm{W}}$           \\
        \hline
        Optical filter gain of receiver, $T\left( \varphi  \right)$ & $0~{\rm{dB}}$  \\
        \hline
        Concentrator gain of receiver, $G\left( \varphi  \right)$ & $0~{\rm{dB}}$  \\
        \hline
        Noise PSD, ${\sigma ^2}$ & $10^{-18}~\rm{A^2/Hz}$   \\
        \hline
        Modulation scheme & 4-QAM   \\
        \hline
        Bandwidth of each subcarrier, ${W}$ & $1~{\rm{MHz}}$   \\
        \hline
    \end{tabular}
\end{table}


\begin{table}[!htbp]
    \centering
    \caption{The tightness of the original constraints \eqref{x_k_idc}, \eqref{dimming_aco}, \eqref{electrical_aco} and \eqref{PA_con3} for the solution of problem \eqref{aco_form}.}\label{compare_cons}
    \begin{tabular}{|l|l|l|}
        \hline
        \diagbox{Definition} {Value} {Scenarios} & \tabincell{l}{${P_o}=0.5$ W,\\${P_e}=20$ W } & \tabincell{l}{${P_o}=0.8$ W,\\${P_e}=10$ W} \\
        \hline
        ${ I_{{\rm{dc}}}^{{\rm{non-clipping}}}}$ & 0.4991 & 0.5491\\
        \hline
        $\min \left( {{x_k}} \right)$ & $-$0.4052 & $-$0.4175\\
        \hline
        $\mathbb{E}\left\{ {{x_{{\rm{dc}},k}}} \right\}$ & 0.4991 W & 0.5491 W\\
        \hline
        $\sum_{k = 0}^{2N - 1} {\mathbb{E}\left\{ {x_{{\rm{dc}},k}^2} \right\}} $ & 8.2369 W & 9.9698 W\\ 
        \hline
    \end{tabular}
\end{table}

In order to solve problem \eqref{aco_form},
we use inequality to deal with constraints \eqref{x_k_idc}, \eqref{dimming_aco}, \eqref{electrical_aco} and \eqref{PA_con3}
into \eqref{power_con1}, \eqref{power_con2} and \eqref{power_con3}.
To verify the validity of the inequality treated constraints,
we substitute the obtained solution of problem \eqref{finite_lowerbound_ACO} into the original constraints \eqref{x_k_idc}, \eqref{dimming_aco} and \eqref{electrical_aco}.
The tightness of the original constraints \eqref{x_k_idc}, \eqref{dimming_aco}, \eqref{electrical_aco} and \eqref{PA_con3} for the solutions of problem \eqref{aco_form} is shown in Table \ref{compare_cons}.
It can be seen that in the scenario of ${P_o}=0.5$ W and ${P_e}=20$ W,
$\mathbb{E}\left\{ {{x_{{\rm{dc}},k}}} \right\}=0.4991$ W, thus constraint \eqref{dimming_aco}, i.e., $\mathbb{E}\left\{ {x_{{\rm{dc}},k}} \right\} \le {P_o}$, is almost tight,
while in the scenario of ${P_o}=0.8$ W and ${P_e}=10$ W,
$\sum\limits_{k = 0}^{2N - 1} {\mathbb{E}\left\{ {x_{{\rm{dc}},k}^2} \right\}}=9.9698$ W,
thus constraint \eqref{electrical_aco}, i.e., $\sum\limits_{k = 0}^{2N - 1} {\mathbb{E}\left\{ {x_{{\rm{dc}},k}^2} \right\}}  \le {P_e}$, is almost tight.
Therefore, our treatment of constraints is valid and the obtained optimal solutions are also high-quality solutions for original problems.

\subsection{Simulation Results of SE Maximization Problems}
In this subsection, we present the results of the proposed power allocation schemes for  maximizing the SE  for
finite-alphabet inputs and lower bound of  the mutual information.



\begin{figure}[htbp]
    \centering
    \begin{minipage}[t]{0.45\textwidth}
        \centering
        \includegraphics[width=\textwidth]{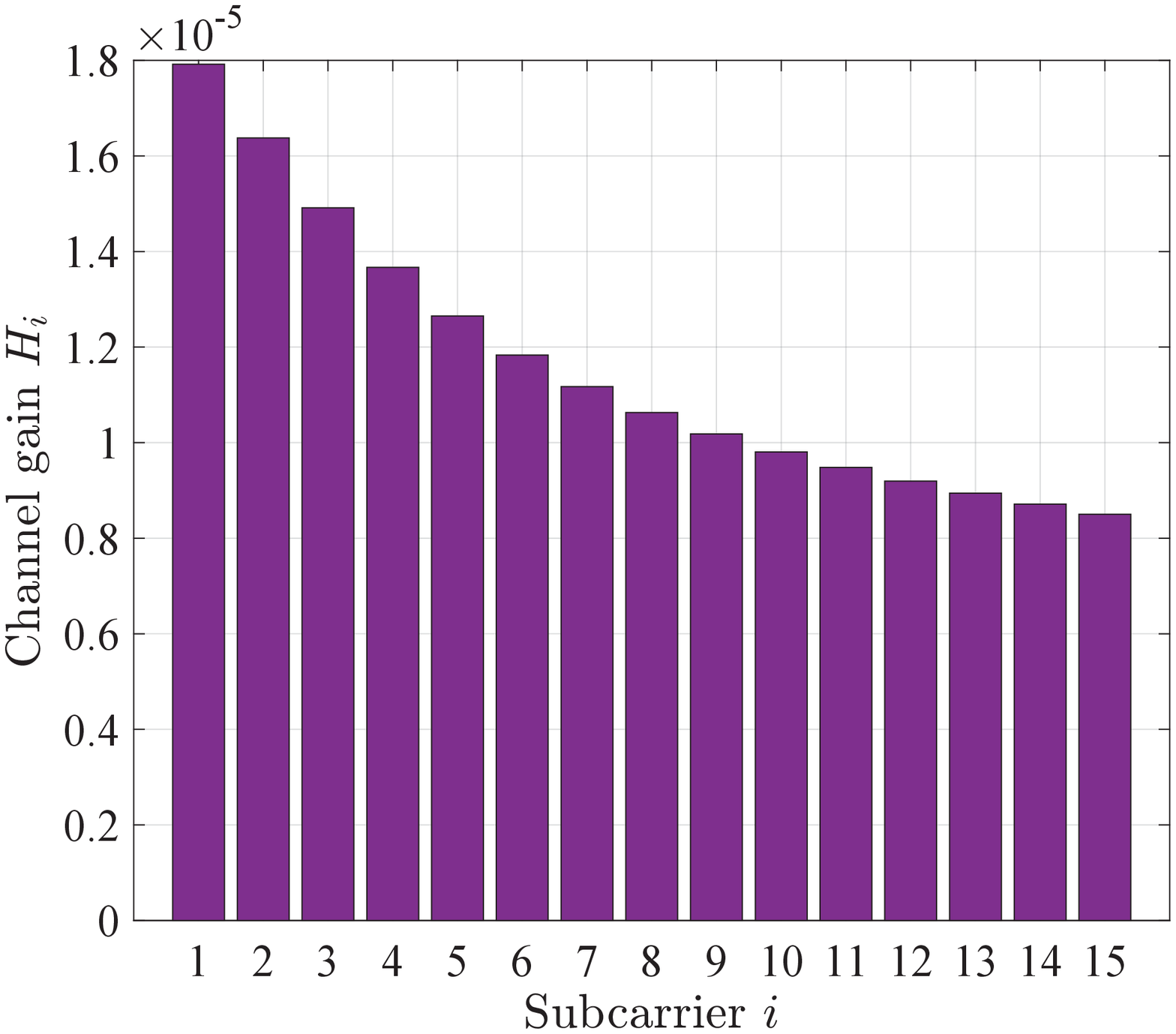}
        \vskip-0.2cm\centering {\footnotesize (a)}
    \end{minipage}
    \begin{minipage}[t]{0.45\textwidth}
        \centering
        \includegraphics[width=\textwidth]{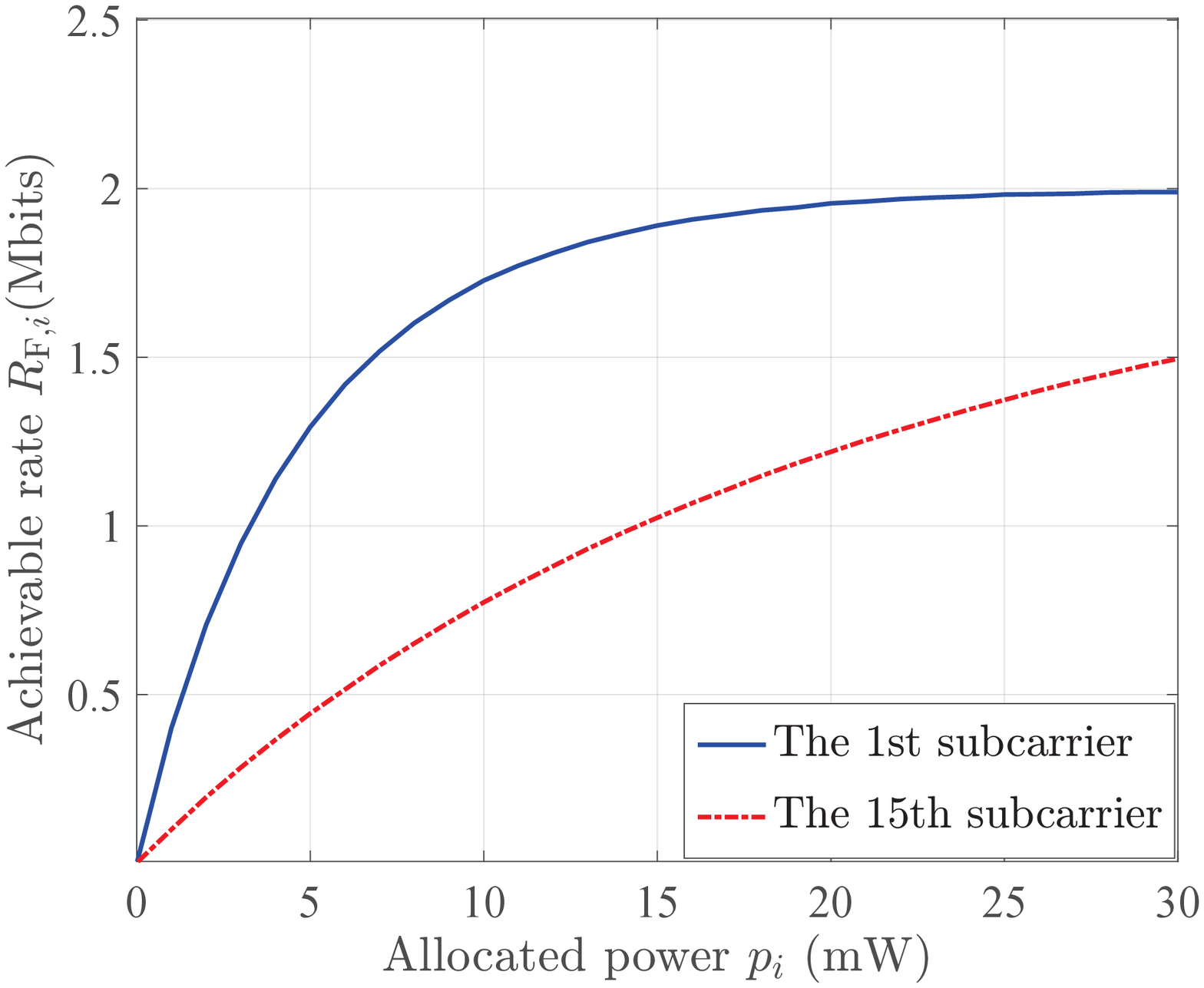}
        \vskip-0.2cm\centering {\footnotesize (b)}
    \end{minipage}
    \begin{minipage}[t]{0.45\textwidth}
        \centering
        \includegraphics[width=\textwidth]{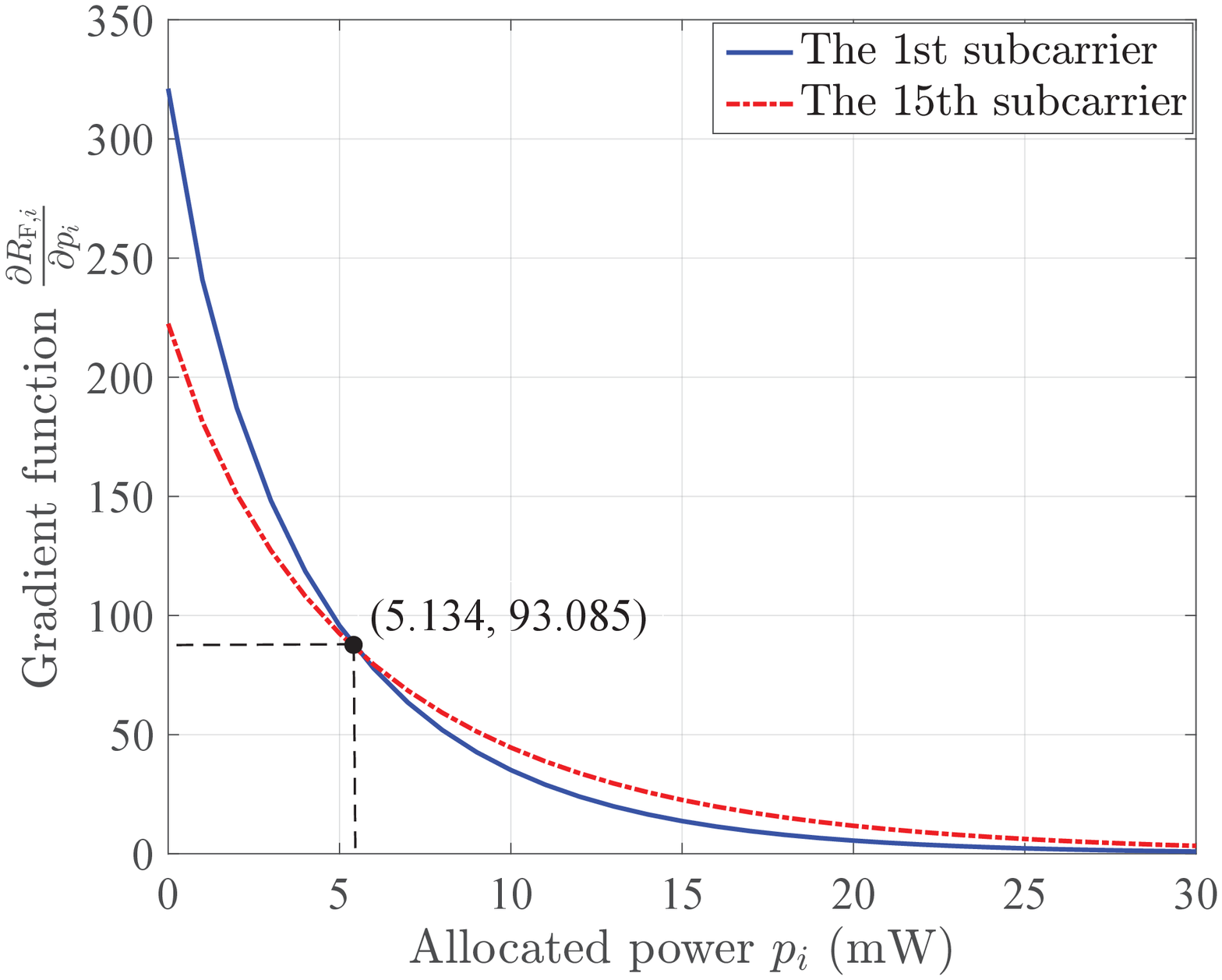}
        \vskip-0.2cm\centering {\footnotesize (c)}
    \end{minipage}
    
    \caption{(a)~Channel gain $H_{i}$ of subcarrier $i$;(b)~Achievable rate ${R_{{\rm{F}},i}}$ versus the allocated power ${p_{i}}$;
        (c)~Gradient function $\frac{{\partial {R_{{\rm{F}},i}}}}{{\partial {p_i}}}$  versus the allocated power ${p_{i}}$.}
    \label{fig_channel_gain}
    \label{fig_Ri_diff_Ri_pi} 
\end{figure}

In order to illustrate the effect of  the difference between channel gain of each subcarrier on power allocation,
the channel gain $H_{i}$, i.e., equation \eqref{channel_model}, of half subcarriers is shown in Fig. \ref{fig_channel_gain} (a).
It can be seen from Fig. \ref{fig_channel_gain} (a) that the channel model in our DCO-OFDM system also has low-pass characteristics.
As the subcarrier index $i$ increases, the corresponding channel gain $H_{i}$ decreases.
The reason is that the channel gain  $H_{i}$ varies over the subcarrier in the dispersive channel model
and high-frequency subcarriers correspond to higher subcarrier index $i$.

%

Fig. \ref{fig_Ri_diff_Ri_pi} (b) shows the achievable rate ${R_{{\rm{F}},i}}$, i.e., equation \eqref{mutual_info_new},
versus the allocated power ${p_{i}}$, in the case of $i=1$ and $i=15$.
It can be seen that as the allocated power ${p_{i}}$ increases,
the achievable rate ${R_{{\rm{F}},1}}$ and ${R_{{\rm{F}},15}}$ first increase fast and then increase slowly,
and ${R_{{\rm{F}},1}}$ gradually approaches ${\log _2}M$ ($M=4$).
This is because the mutual information of $M$-ary discrete constellation modulation can not exceed ${\log _2}M$.
Moreover, ${R_{{\rm{F}},1}}$ is higher than ${R_{{\rm{F}},15}}$ since ${H_1} > {H_{15}}$.

Fig. \ref{fig_Ri_diff_Ri_pi} (c) shows  the gradient function $\frac{{\partial {R_{{\rm{F}},i}}}}{{\partial {p_{i}}}}$,
i.e., equation \eqref{MMSE_MI_aco_2}, versus the allocated power ${p_{i}}$, in the case of $i=1$ and $i=15$.
It can be seen that both $\frac{{\partial {R_{{\rm{F}},1}}}}{{\partial {p_{1}}}}$ and $\frac{{\partial {R_{{\rm{F}},15}}}}{{\partial {p_{15}}}}$
decrease with the increase of ${p_{i}}$, and eventually tend to $0$.
When ${p_{i}} \le 5.134$ mW, $\frac{{\partial {R_{{\rm{F}},1}}}}{{\partial {p_{1}}}} \ge \frac{{\partial {R_{{\rm{F}},15}}}}{{\partial {p_{15}}}}$;
when ${p_{i}} > 5.134$ mW, $\frac{{\partial {R_{{\rm{F}},1}}}}{{\partial {p_{1}}}} < \frac{{\partial {R_{{\rm{F}},15}}}}{{\partial {p_{15}}}}$.
Fig. \ref{fig_Ri_diff_Ri_pi} (b) illustrates that, compared with subcarriers with small channel gains,
when the allocated power is small, the subcarriers with large channel gains have larger gradient of rate.
When the allocated power is large, the subcarriers with large channel gains have smaller gradient of rate.

\begin{figure}[htbp]
  \centering
        \begin{minipage}[t]{0.45\textwidth}
             \centering
             \includegraphics[width=\textwidth]{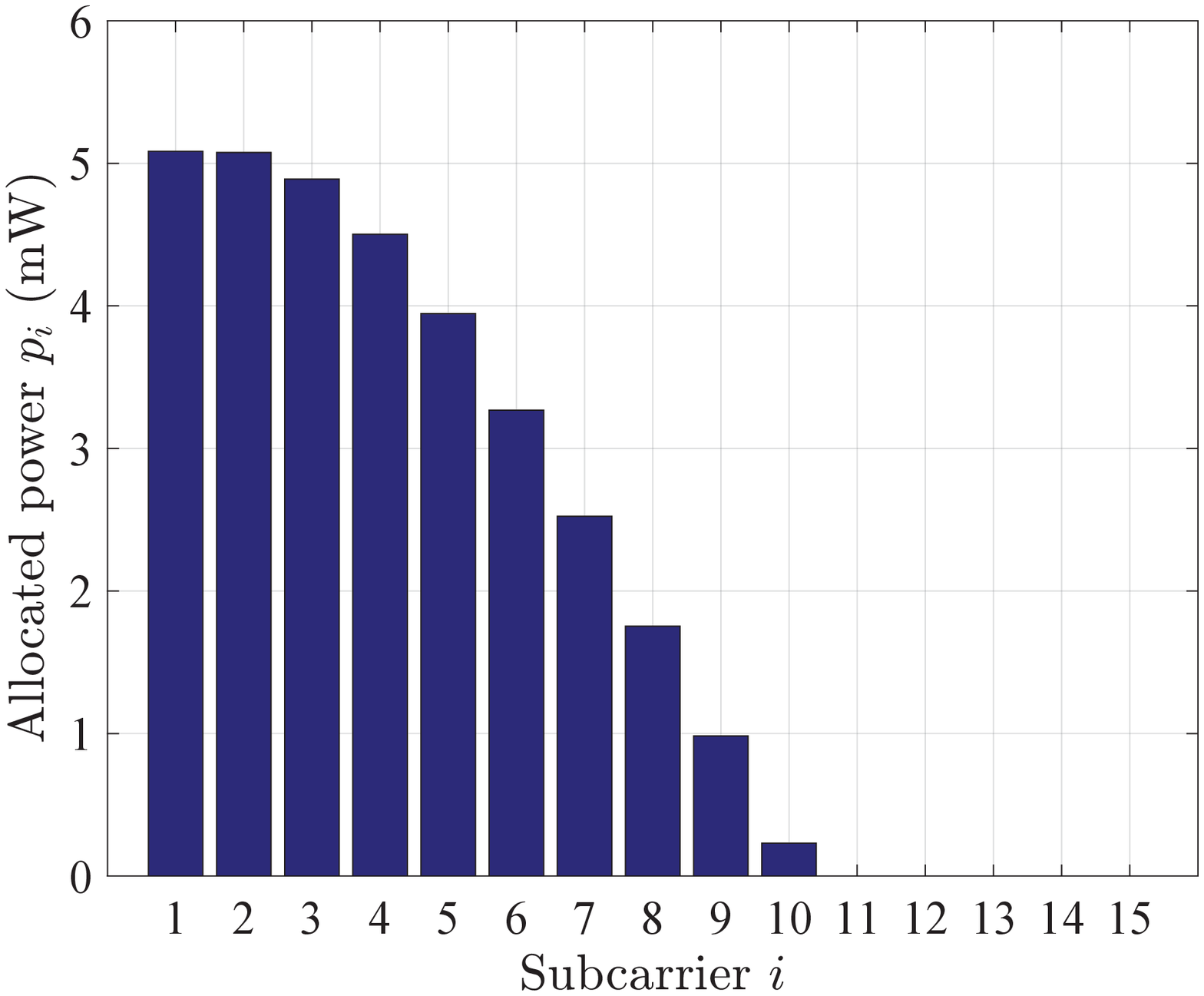}
             \vskip-0.2cm\centering {\footnotesize (a)}
        \end{minipage}
        \begin{minipage}[t]{0.45\textwidth}
             \centering
             \includegraphics[width=\textwidth]{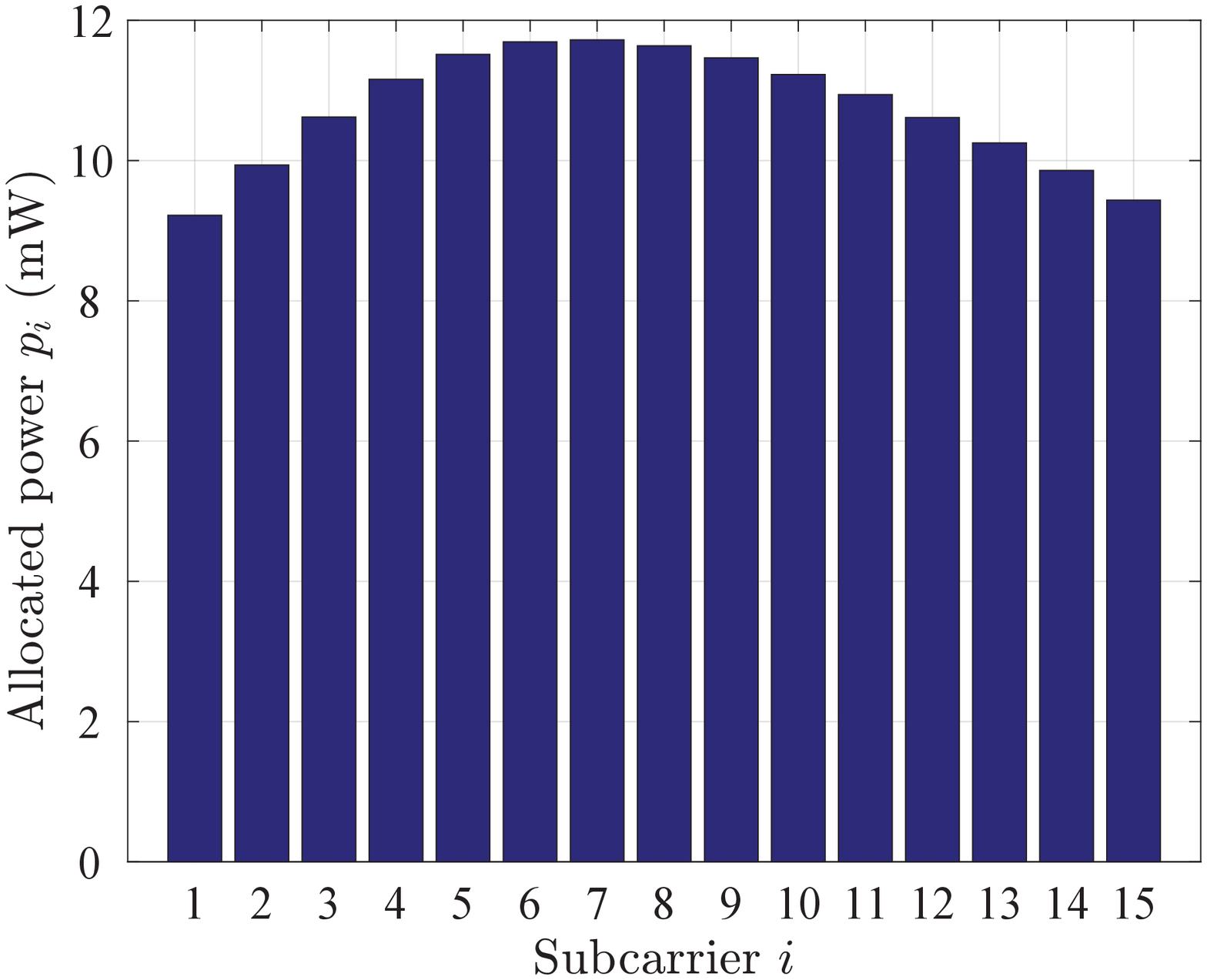}
             \vskip-0.2cm\centering {\footnotesize (b)}
        \end{minipage}
        \begin{minipage}[t]{0.45\textwidth}
             \centering
             \includegraphics[width=\textwidth]{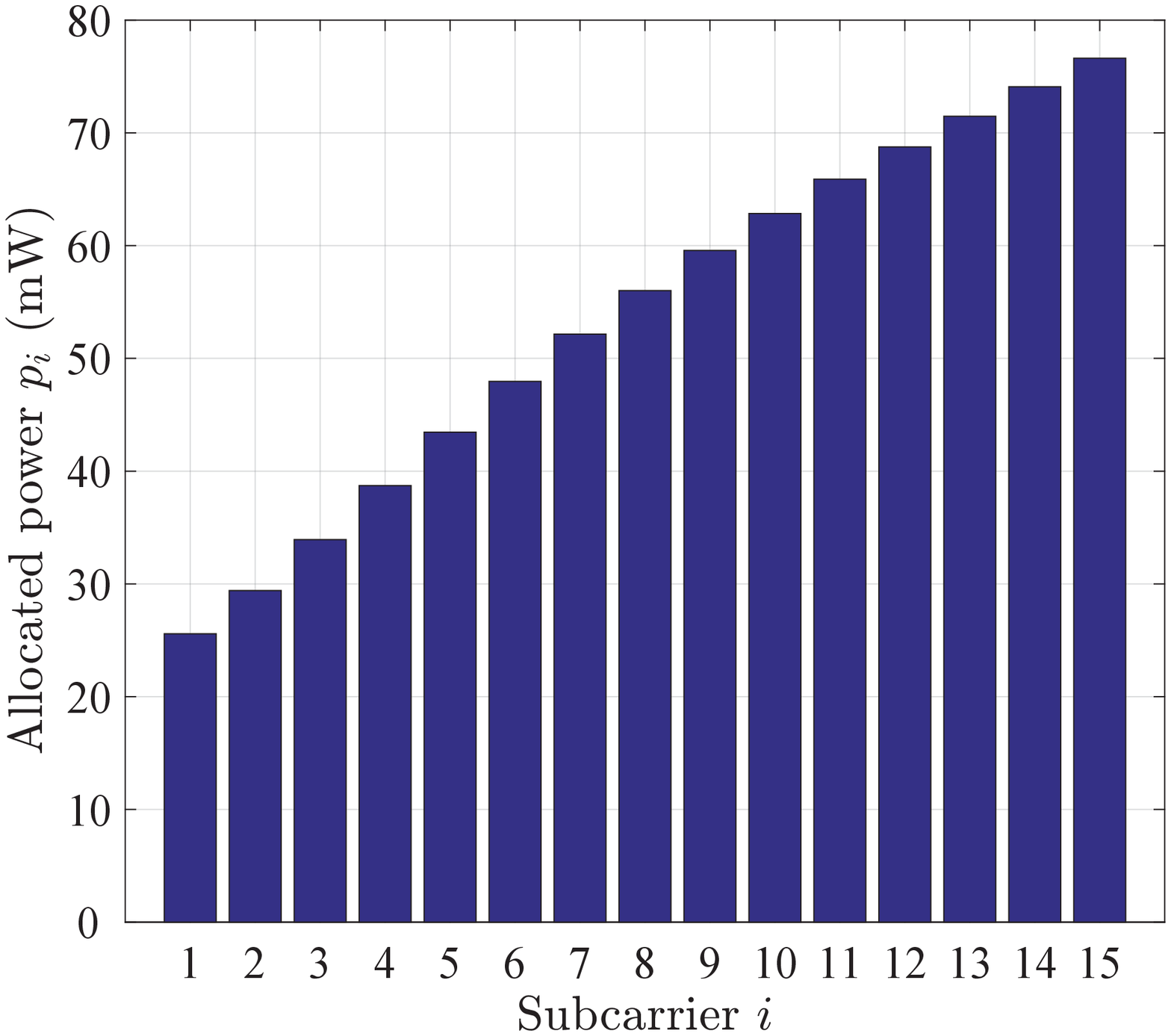}
             \vskip-0.2cm\centering {\footnotesize (c)}
        \end{minipage}
 \caption{Allocated power $p_{i}$ of  subcarrier $i$ of $\text{SE}_\text{F}$ with ${P_o} = 10$ W (a)~${P_e} = 2$ W; (b)~${P_e} = 10$ W; (c)~${P_e} = 50$ W.}
  \label{fig_allocated_pi_mmse} 
\end{figure}

%


In the following, three different power allocation scenarios for the SE maximization problem \eqref{aco_finite} are compared
in the case of low, medium, and high total {transmitted} power.
Specially, Fig. \ref{fig_allocated_pi_mmse} (a), (b) and (c) are achieved with the same average optical power ${P_o} = 10$ W,
and different total electrical {transmitted} power ${P_e} = 2$ W, ${P_e} = 10$ W and ${P_e} = 50$ W respectively.


It can be seen from Fig. \ref{fig_allocated_pi_mmse} (a) that when the total allocated power is low (${P_e} = 2$ W),
more power is allocated to subcarriers with larger channel gains,
which is similar to the classical water-filling solution with the Gaussian distribution inputs.
Combining Fig. \ref{fig_allocated_pi_mmse} (a) and Fig. \ref{fig_Ri_diff_Ri_pi},
it can be seen that this is because when the  allocated power is small,
the rate of the subcarriers with larger channel gains increases faster than the subcarriers with smaller channel gains,
thus more power should be allocated to the subcarriers with larger gradient of rate to maximize the $\text{SE}_\text{F}$ of the system.

Fig. \ref{fig_allocated_pi_mmse} (b) shows that for the case of medium total allocated power (${P_e} = 10$ W) , more power is allocated to the subcarriers with moderate channel gains.
Combining Fig. \ref{fig_allocated_pi_mmse} (b) and Fig. \ref{fig_Ri_diff_Ri_pi},
it can be seen that this is because as the allocated power increases,
the rate of subcarriers with smaller channel gains increases faster than the rate of subcarriers with larger channel gains at this time.
Therefore, more power is allocated to the subcarriers with smaller channel gains to maximize the $\text{SE}_\text{F}$ of the system,
i.e., power is always preferentially allocated to the subcarrier with the largest gradient of rate.

It can be seen from Fig. \ref{fig_allocated_pi_mmse} (c) that when the total allocated power is high (${P_e} = 50$ W),
more power is allocated to subcarriers with smaller channel gains,
which is exactly opposite with the water-filling policy.
Combining Fig. \ref{fig_allocated_pi_mmse} (c) and Fig. \ref{fig_Ri_diff_Ri_pi},
it can be seen that in this case, the rate of subcarriers with larger channel gains tends to be saturated (${\log _2}M$),
thus there is little incentive to allocate further power to such subcarriers.
Rather, to maximize the $\text{SE}_\text{F}$ of the system,
the additional power is better allocated to subcarriers with smaller channel gains whose rate is still far from saturation.

\begin{figure}[htbp]
    \centering
    \begin{minipage}[t]{0.45\textwidth}
        \centering
        \includegraphics[width=\textwidth]{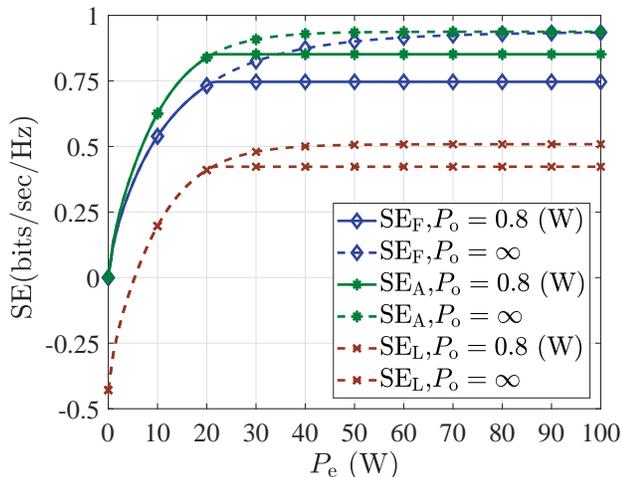}
        \vskip-0.2cm\centering {\footnotesize (a)}
    \end{minipage}
    \begin{minipage}[t]{0.45\textwidth}
        \centering
        \includegraphics[width=\textwidth]{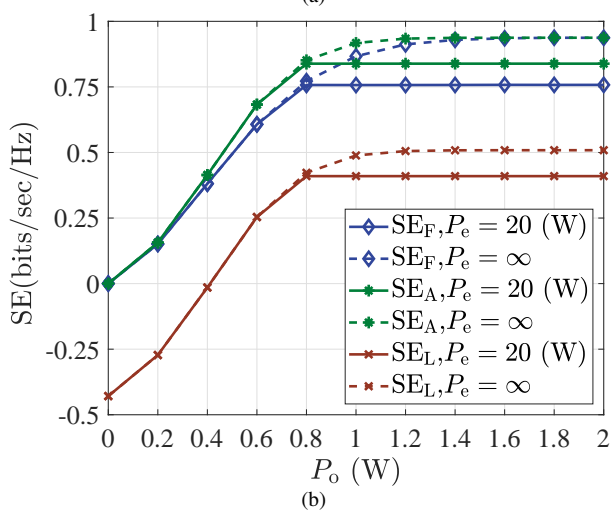}
        \vskip-0.2cm\centering {\footnotesize (b)}
    \end{minipage}
    \caption{(a)~$\text{SE}_\text{F}$, $\text{SE}_{\text{A}}$ and $\text{SE}_\text{L}$  versus electrical power budget $P_{\mathrm{e}}$ with two different optical power budget $P_{\mathrm{o}} = 0.8$ W and $P_{\mathrm{o}} = \infty$; (b)~$\text{SE}_\text{F}$, $\text{SE}_{\text{A}}$ and $\text{SE}_\text{L}$ versus optical power budget ${P_{o}}$ with two different electrical power budget ${P_e} = 20$ W and ${P_e} = \infty$.}
    \label{fig_SE_pe}
    \label{fig_SE_po}
\end{figure}

Fig. \ref{fig_SE_pe} (a) illustrates $\text{SE}_\text{F}${, $\text{SE}_\text{A}$} and $\text{SE}_\text{L}$  versus total electrical {transmitted} power budget $P_{\mathrm{e}}$
with two different optical power budget $P_{\mathrm{o}} = 0.8$ W and $P_{\mathrm{o}} = \infty$ (without optical power constraint), respectively.
For the case of $P_{\mathrm{o}} = 0.8$ W, as $P_{\mathrm{e}}$ increases, $\text{SE}_\text{F}$, $\text{SE}_\text{A}$ and $\text{SE}_\text{L}$ first increase and then get restricted into a constant.
The reason is that the total allocated power is restricted by the optical power $P_{\mathrm{o}} = 0.8$ W.
For the case of $P_{\mathrm{o}} = \infty$, as $P_{\mathrm{e}}$ increases, $\text{SE}_\text{F}${, $\text{SE}_\text{A}$} and $\text{SE}_\text{L}$  first increase  and then remain constant.
This is because the mutual information of $M$-ary discrete constellation modulation  can not exceed ${\log _2}M$.
Moreover, similar to {$\text{SE}_\text{A}$ and} $\text{SE}_\text{L}$, when $P_{\mathrm{e}} > 20$ W, $\text{SE}_\text{F}$ with $P_{\mathrm{o}} = \infty$ is higher than $\text{SE}_\text{F}$ with $P_{\mathrm{o}} = 0.8$ W.
This is because the total allocated power is restricted when $P_{\mathrm{o}}$ takes the value of $0.8$ W.
It can be seen that, using the optimal power allocation policy proposed in this paper,
$\text{SE}_\text{F}$ is significantly higher than $\text{SE}_\text{L}$
since ${{R_{{\rm{L,total}}}}\left( {\left\{ {{p_i}} \right\}} \right)}$ is the lower bound of ${{R_{{\rm{F,total}}}}\left( {\left\{ {{p_i}} \right\}} \right)}$. {Meanwhile, $\text{SE}_\text{F}$ is slightly lower than $\text{SE}_\text{A}$ in the medium $P_{\mathrm{e}}$, however, they will tend to coincide when $P_{\mathrm{e}}\to\infty$ and $P_{\mathrm{e}}\to 0$, which consistent with the feature of ${{R_{{\rm{A,total}}}}\left( {\left\{ {{p_i}} \right\}} \right)}$ [36].}

{Fig. \ref{fig_SE_po} (b)} depicts $\text{SE}_\text{F}${, $\text{SE}_\text{A}$} and $\text{SE}_\text{L}$ versus average optical power budget ${P_{o}}$ with two different electrical power budgets ${P_e} = 20$ W and ${P_e} = \infty$ (without electrical power constraint), respectively.
For the case of ${P_e} = 20$ W, as ${P_{o}}$ increases, $\text{SE}_\text{F}${, $\text{SE}_\text{A}$} and $\text{SE}_\text{L}$ first increase and then get restricted as a constant.
This is because the total allocated power is limited by the electrical power budget ${P_e} = 20$ W.
When ${P_e} = \infty$, as ${P_{o}}$ increases, $\text{SE}_\text{F}${, $\text{SE}_\text{A}$} and $\text{SE}_\text{L}$  first increase  and then remain constant.
In addition, similar to {$\text{SE}_\text{A}$ and} $\text{SE}_\text{L}$, when ${P_{o}} > 0.6$ W, $\text{SE}_\text{F}$ with ${P_e} = \infty$ is higher than $\text{SE}_\text{F}$ with ${P_e} = 20$ W.
This is because the total allocated power is restricted when ${P_e}$ takes the value of $20$ W.
Besides, we can observe that $\text{SE}_\text{F}$ is significantly higher than $\text{SE}_\text{L}${, and $\text{SE}_\text{A}$ is slightly highter than $\text{SE}_\text{A}$ but converges if $P_{\mathrm{e}}\to\infty$ and $P_{\mathrm{e}}\to 0$}.

\subsection{Simulation Results of EE Maximization Problems}
In this subsection, we present the simulation results of the proposed power allocation schemes for  maximizing the EE  for
finite-alphabet inputs and lower bound of  the mutual information.

\begin{figure}[htbp]
  \centering
        \begin{minipage}[t]{0.45\textwidth}
             \centering
             \includegraphics[width=\textwidth]{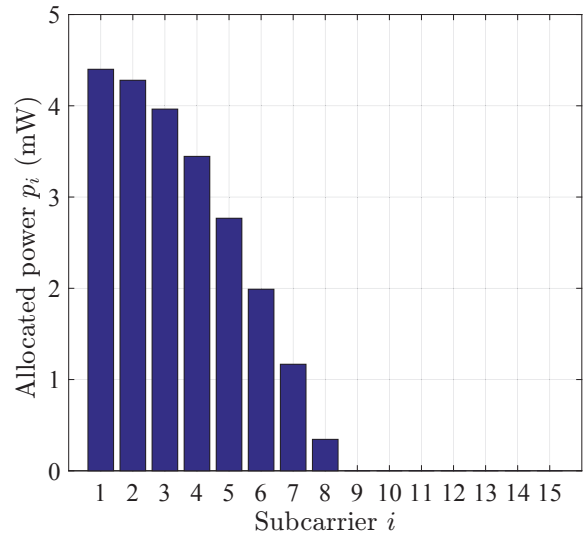}
             \vskip-0.2cm\centering {\footnotesize (a)}
        \end{minipage}
        \begin{minipage}[t]{0.45\textwidth}
             \centering
             \includegraphics[width=\textwidth]{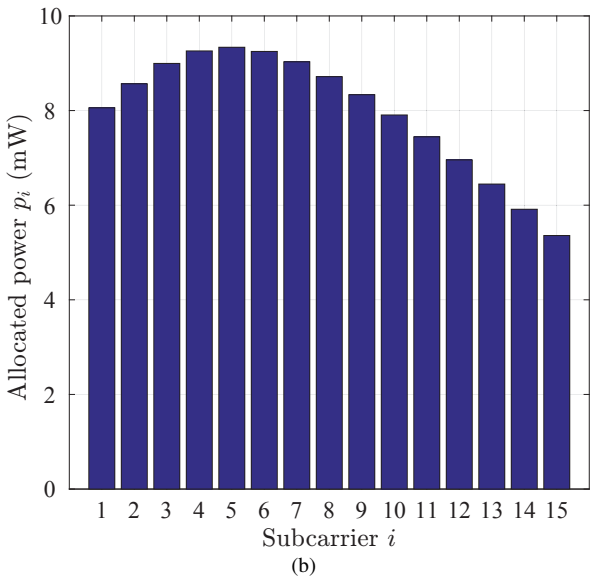}
             \vskip-0.2cm\centering {\footnotesize (b)}
        \end{minipage}
        \begin{minipage}[t]{0.45\textwidth}
             \centering
             \includegraphics[width=\textwidth]{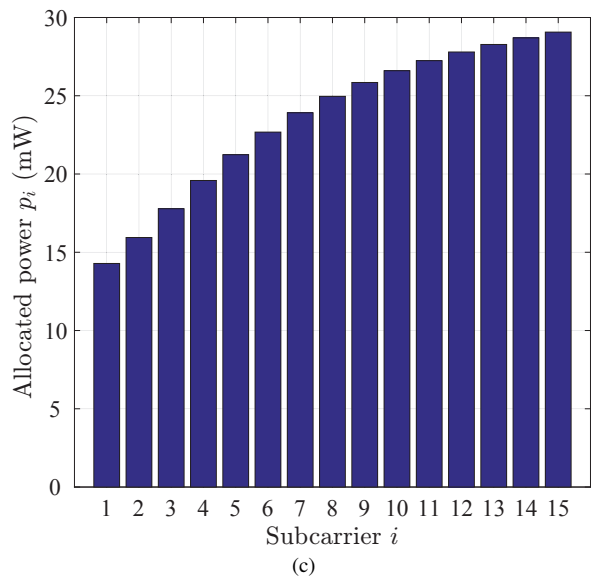}
             \vskip-0.2cm\centering {\footnotesize (c)}
        \end{minipage}
 \caption{Allocated power $p_{i}$ of  subcarrier $i$ of $\text{EE}_\text{F}$ with ${P_o} = 1$ W, ${P_e} = 22$ W,
 (a)~ $\gamma= 0.156$ Mbits/sec/Hz ($\bar R = 5$ Mbits/sec);  
 (b)~ $\gamma= 0.469$ Mbits/sec/Hz ($\bar R = 15$ Mbits/sec);  
 (c)~ $\gamma= 0.781$ Mbits/sec/Hz ($\bar R = 25$ Mbits/sec).}  
  \label{fig_allocated_EE_pi} 
\end{figure}

%

To illustrate the three scenarios when the SE threshold $\gamma$ is small, medium and large,
the allocated power $p_{i}$ of  subcarrier $i$ of $\text{EE}_\text{F}$ with $\gamma= 0.156$ Mbits/sec/Hz ($\bar R = 5$ Mbits/sec),
$\gamma= 0.469$ Mbits/sec/Hz ($\bar R = 15$ Mbits/sec) and $\gamma= 0.781$ Mbits/sec/Hz ($\bar R = 25$ Mbits/sec), where ${P_o} = 1$ W, ${P_e} = 22$ W,
are shown in Fig. \ref{fig_allocated_EE_pi} (a), (b) and (c) respectively.

{It can be seen from Fig. \ref{fig_allocated_EE_pi} (a) that when the SE threshold $\gamma$ is small ($\gamma= 0.156$ Mbits/sec/Hz), more power is allocated to subcarriers with larger channel gains, and the required power increases with the improvement of the SE threshold $\gamma$. Besides, Fig. \ref{fig_Ri_diff_Ri_pi} (c) shows that the gradient of the rate increases synchronously with the channel gain. Thus, more power would be allocated to the subcarriers with a larger gradient of the rate to maximize the $\text{EE}_\text{F}$ of the system.}

Fig. \ref{fig_allocated_EE_pi} (b) shows that for the case of medium SE threshold $\gamma$ ($\gamma= 0.469$ Mbits/sec/Hz),
more power is allocated to the subcarriers with moderate channel gains,
and as the SE threshold $\gamma$ increases, the power that needs to be allocated becomes larger. Combining with Fig. \ref{fig_Ri_diff_Ri_pi} (c),  it can be seen that to maximize the $\text{EE}_\text{F}$ of the system, the power is always preferentially allocated to the subcarrier with the largest gradient of the rate.

It can be seen from Fig. \ref{fig_allocated_EE_pi} (c) that when the SE threshold $\gamma$ is large ($\gamma= 0.781$ Mbits/sec/Hz),
more power is allocated to subcarriers with smaller channel gains.
Combining Fig. \ref{fig_allocated_EE_pi} (c) and Fig. \ref{fig_Ri_diff_Ri_pi} (c),
it can be seen that when the allocated power $p_{i}$ is large,
to achieve the same gradient of the rate, the power allocated to subcarriers with larger channel gains
is much greater than the power allocated to subcarriers with smaller channel gains.
Thus, the additional power is preferentially allocated to subcarriers with smaller channel gains to maximize the $\text{EE}_\text{F}$ of the system.

\begin{figure}[htbp]
  \centering
  \begin{minipage}[t]{0.45\textwidth}
       \centering
       \includegraphics[width=\textwidth]{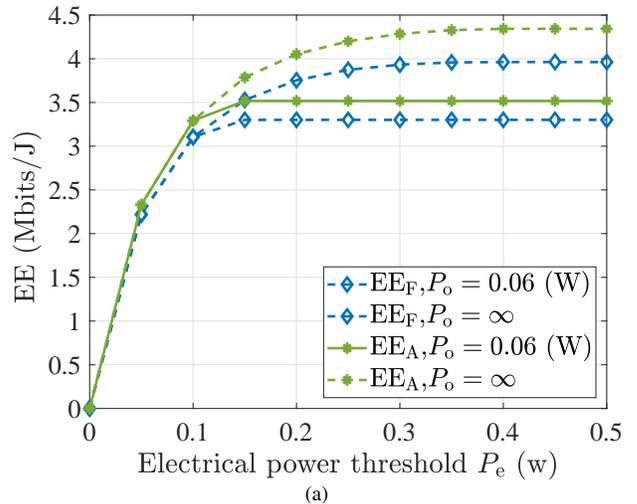}
       \vskip-0.2cm\centering {\footnotesize (a)}
  \end{minipage}
  \begin{minipage}[t]{0.45\textwidth}
       \centering
       \includegraphics[width=\textwidth]{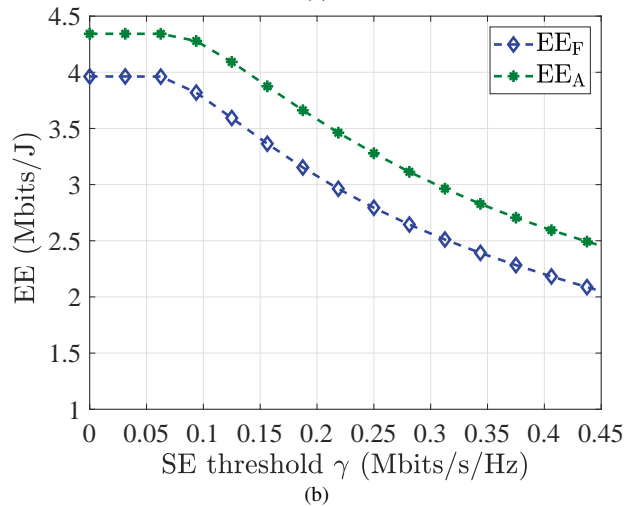}
       \vskip-0.2cm\centering {\footnotesize (b)}
  \end{minipage}
\caption{(a)~${\text{EE}}_\text{F}$ and ${\text{EE}_\text{A}}$ versus electrical power budget ${P_e}$ with SE threshold $\gamma= 0.009$ Mbits/sec/Hz ($\bar R = 0.3$ Mbits/sec) and two different optical power budget  $P_{\mathrm{o}} = 0.06$ W  and  $P_{\mathrm{o}} = \infty$; (b)~${\text{EE}}_\text{F}$, ${\text{EE}_\text{A}}$ and ${\text{EE}_\text{L}}$ versus SE threshold $\gamma$  with  electrical power budget ${P_{\mathrm{e}}} = 5$ W and optical power budget $P_{\mathrm{o}} = 1$ W.}
\label{fig_EE_pe}
\label{fig_EE_rate}
\end{figure}

Fig. \ref{fig_EE_pe} (a) illustrates $\text{EE}_\text{F}$ and {$\text{EE}_\text{A}$}  versus electrical power budget $P_{\mathrm{e}}$
with SE constraint $\gamma= 0.009$ Mbits/sec/Hz ($\bar R = 0.3$ Mbits/sec) and two different optical power budget $P_{\mathrm{o}} = 0.06$ W and $P_{\mathrm{o}} = \infty$ (without optical power constraint), respectively.
For the case of $P_{\mathrm{o}} = 0.06$ W, as ${P_{\mathrm{e}}}$ increases, $\text{EE}_\text{F}$ and {$\text{EE}_\text{A}$} first increase and then get restricted into a constant.
The reason is that the total allocated power is restricted by the optical power $P_{\mathrm{o}} = 0.06$ W.
For the case of $P_{\mathrm{o}} = \infty$, as $P_{\mathrm{e}}$ increases, $\text{EE}_\text{F}$ and {$\text{EE}_\text{A}$}  first increase  and then remain constant.
This is because $\text{EE}_\text{F}$ and {$\text{EE}_\text{A}$} remain constant when they reach their maximum values respectively.
Moreover, similar to {$\text{EE}_\text{A}$}, for the large $P_{\mathrm{e}}$, $\text{EE}_\text{F}$ with $P_{\mathrm{o}} = \infty$ is higher than $\text{SE}_\text{F}$ with $P_{\mathrm{o}} = 0.06$ W
since $\text{EE}_\text{L}$ and $\text{EE}_\text{F}$ are restricted when $P_{\mathrm{o}}$ takes the value of $0.06$ W.
It can be seen that $\text{EE}_\text{F}$ is  {lower} than {$\text{EE}_\text{A}$}.

Fig. \ref{fig_EE_rate} (b) depicts ${\text{EE}}_\text{F}$ and { $\text{EE}_\text{A}$} versus SE threshold $\gamma$
with electrical power budget $P_{\mathrm{e}} = 5$ W and optical power budget $P_{\mathrm{o}} = 1$ W.
We observe that ${\text{EE}}_\text{F}$ and { $\text{EE}_\text{A}$} both first keep stable and then decrease as SE threshold $\gamma$ increases.
This is because when the SE threshold $\gamma$ is small,
the performed power allocation can easily satisfy the SE requirement  and thus EE keeps as a constant.
When the SE threshold $\gamma$ becomes larger,
more power needs to be consumed to satisfy the rate constraint, therefore the EE decreases.

{
    \emph{C. Simulation Results of Computational Complexity}
    
    In this subsection, we present the average executing time versus half of subcarrier number $N$ to evaluate the computational complexity of the proposed power allocation schemes for SE- and EE-maximization problem.
    For given $N$ and other fixed parameters, the average executing time is calculated according on $10000$ repeated simulations, which are performed by MATLAB (2020b) with Intel(R) Core(TM) i9-10900K 3.70 GHz CPU and 32 GB RAM.
    
        \begin{table}[htbp]
        \centering
        \caption{The average executing time (ms) of different scheme versus half of subcarrier number $N$.}\label{ave_exe_time}
        \begin{tabular}{|c|c|c||c|c|}
            \hline
            \diagbox{$N$} {Time} {Scheme} & $\mathrm{SE_F}$ & {$\mathrm{SE_A}$} & $\mathrm{EE_F}$ & {$\mathrm{EE_A}$}\\
            \hline
            4  & $8.34 $   & $2.91$ & $18220$ & $1030$ \\
            \hline
            8  & $48.77  $ & $2.83 $ & $27630$ & $1040$ \\
            \hline
            16 & $126.51 $ & $2.99 $ & $72500$ & $1040$ \\
            \hline
        \end{tabular}
    \end{table}
    
    For the SE-maximization problem, through the mercury-water-filling method, Table \ref{ave_exe_time} depicts the comparing of the scheme base on $R_{\mathrm{F,total}}\left(\left\{p_i\right\}\right) $ and based on { $R_{\mathrm{A,total}}\left(\left\{p_i\right\}\right) $} with electrical power budget ${P_e} = 20$ W and optical power budget ${P_o} = 5$ W. As shows, the CPU time of the proposed  schemes based on $R_{\mathrm{F,total}}\left(\left\{p_i\right\}\right)$ increases as $N$ increases, however, for the proposed  schemes based on { $R_{\mathrm{A,total}}\left(\left\{p_i\right\}\right) $}, it almost is a constant. Meanwhile, it is obvious that the scheme based on { $R_{\mathrm{A,total}}\left(\left\{p_i\right\}\right) $} is much faster than the scheme based on $R_{\mathrm{F,total}}\left(\left\{p_i\right\}\right)$.
    
    For the EE-maximization problem, Table \ref{ave_exe_time} presents the difference of the scheme based on $R_{\mathrm{F,total}}\left(\left\{p_i\right\}\right) $ and  { $R_{\mathrm{A,total}}\left(\left\{p_i\right\}\right) $} with  electrical power budget ${P_e} = 20$ W, optical power budget ${P_o} = 5$ W, and SE threshold $\gamma= 0.009$ Mbits/sec/Hz,and we can draw a conclusion as same as the comparing between $\mathrm{SE_F}$ and $\mathrm{SE_L}$.

}

\section{Conclusion}
 In this work, we investigated the
{bound of the information transmission rate}
  of {the} DCO-OFDM system with finite-alphabet inputs,
and   proposed the optimal power allocation schemes to achieve maximum SE and maximum EE of {the} DCO-OFDM system, respectively.
We first derived the exact achievable rate expression  without information loss.
Then, we developed the power allocation to maximize SE   based on the lower bound of achievable rate.
 Furthermore,  we exploited the  KKT conditions and the relationship between the mutual information and MMSE,
and derived the  multi-level mercury-water-filling  power allocation scheme for SE maximization.
Moreover, we proposed the Dinkelbach-type power allocation {scheme and} obtained the optimal power allocation  for EE maximization.
Finally, numerical results revealed that the proposed multi-level mercury-water-filling scheme of SE maximization
based on the exact mutual information {depends} on  both channel gain of each subcarrier and  the total {transmitted power} constraint.
When   the total {transmitted power}  is high, the power allocation of each  subcarrier is inversely proportional to the channel gain,
which is  different from  that of  the classical water-filling method.
Besides, we revealed that for the EE maximization problem with finite-alphabet inputs,
 the power allocation of each subcarrier is proportional to the channel gain for {the}    low  SE requirement.
While for the high SE requirement, the power allocation of  each  subcarrier is inversely proportional to the channel gain.
Besides, under the same constraints, the value of SE and EE based on the exact mutual information is always higher than that based on the lower bound.

\begin{appendices}
\section{proof of \eqref{mutual_info_new}}
According to equation \eqref{signal_model},
the conditional probability density function $p\left( {{Y_i}|{X_i}} \right)$ and the probability density function $p\left( {{Y_i}} \right)$
corresponding to the channel output ${Y_i}$ are respectively expressed as
\begin{subequations}
\begin{align}
p\left( {{Y_i}|{X_i}} \right) = \frac{1}{{\pi {\sigma ^2}W}}\exp \left( { - \frac{{{{\left| {{Y_i} - {H_i}\sqrt {{p_i}} {X_i}} \right|}^2}}}{{{\sigma ^2}W}}} \right), \label{con_pdf}\\
p\left( {{Y_i}} \right) = {\mathbb{E}_{{X_i}}}\left\{ {p\left( {{Y_i}|{X_i}} \right)} \right\} = \frac{1}{M}\sum\limits_{k = 1}^M {p\left( {{Y_i}|{X_{i,k}}} \right)} . \label{pdf}
\end{align}
\end{subequations}

Then, the mutual information between the channel input ${X_i}$ and channel output ${Y_i}$ can be obtained as
\begin{subequations}
\begin{align}
&I\left( {{X_i};{Y_i}} \right)  =W\sum\limits_{{X_i}} {\int\limits_{{Y_i}} {p\left( {{X_i},{Y_i}} \right){{\log }_2}\frac{{p\left( {{X_i},{Y_i}} \right)}}{{p\left( {{X_i}} \right)p\left( {{Y_i}} \right)}}\,\mathrm{d} Y_i} } \label{I_derivation_1} \\
 &= W\sum\limits_{n = 1}^M {\int\limits_{{Y_i}} {\frac{1}{M}p\left( {{Y_i}|{X_{i,n}}} \right)} {{\log }_2}\frac{{p\left( {{Y_i}|{X_{i,n}}} \right)}}{{p\left( {{Y_i}} \right)}}\,\mathrm{d} Y_i}
 \label{I_derivation_2}  \\
 &= \frac{W}{M}\sum\limits_{n = 1}^M {\int\limits_{{Y_i}} {p\left( {{Y_i}|{X_{i,n}}} \right)} {{\log }_2}\frac{{p\left( {{Y_i}|{X_{i,n}}} \right)}}{{\sum\limits_{k = 1}^M {\frac{1}{M}p\left( {{Y_i}|{X_{i,k}}} \right)} }}\,\mathrm{d} Y_i} \label{I_derivation_3}  \\
& = \frac{W}{M}\sum\limits_{n = 1}^M {\int\limits_{{Y_i}} {p\left( {{Y_i}|{X_{i,n}}} \right)} {{\log }_2}\frac{{\exp \left( { - \frac{{{{\left| {{Y_i} - {H_i}\sqrt {{p_i}} {X_{i,n}}} \right|}^2}}}{{{\sigma ^2}W}}} \right)}}{{\sum\limits_{k = 1}^M {\frac{1}{M}\exp \left( { - \frac{{{{\left| {{Y_i} - {H_i}\sqrt {{p_i}} {X_{i,k}}} \right|}^2}}}{{{\sigma ^2}W}}} \right)} }}\,\mathrm{d} Y_i}  \label{I_derivation_4}  \\
& =  - \frac{W}{M}\sum\limits_{n = 1}^M {\int\limits_{{Z_i}} {p\left( {{Z_i}} \right)} {{\log }_2}\sum\limits_{k = 1}^M {\frac{1}{M}\exp \left( { - {d_{n,k}} + \frac{{{{\left| {{Z_i}} \right|}^2}}}{{{\sigma ^2}W}}} \right)} \,\mathrm{d} Z_i}  \\
&  =  - \frac{W}{M}\sum\limits_{n = 1}^M {{\mathbb{E}_{{Z_i}}}\left\{ {{{\log }_2}\sum\limits_{k = 1}^M {\frac{1}{M}\exp \left( { - {d_{n,k}} + \frac{{{{\left| {{Z_i}} \right|}^2}}}{{{\sigma ^2}W}}} \right)} } \right\}}  \\
&  =  - \frac{W}{M}\sum\limits_{n = 1}^M {{\mathbb{E}_{{Z_i}}}\left\{ {{{\log }_2}\frac{1}{M}\exp \left( {\frac{{{{\left| {{Z_i}} \right|}^2}}}{{{\sigma ^2}W}}} \right)} \right\}}  \nonumber\\
&~~~~~~~~~~~~~~~- \frac{W}{M}\sum\limits_{n = 1}^M {{\mathbb{E}_{{Z_i}}}\left\{ {{{\log }_2}\sum\limits_{k = 1}^M {\exp \left( { - {d_{n,k}}} \right)} } \right\}}  \\
&=-\frac{W}{M}\sum\limits_{n=1}^{M}\mathbb{E}_{Z_i}\left\{\log_2\left(\frac{\exp\left(\frac{\left|Z_i\right|^2}{\sigma^2 W}\right)\sum\limits_{k=1}^{M}\exp\left(-d_{n,k}\right)}{M}\right)\right\}\\
& = W\left( {{{\log }_2}M - \frac{1}{{\ln 2}}} \right) \nonumber\\
&~~~~~~~~~~~~~- \sum\limits_{n = 1}^M \frac{W}{M}{\mathbb{E}_{{Z_i}}}\left\{ {{{\log }_2}\sum\limits_{k = 1}^M {\exp \left( { - {d_{n,k}}} \right)} } \right\},  \label{E_zi_2}
 \end{align}
 \end{subequations}
where \eqref{I_derivation_3} is due to \eqref{pdf}, \eqref{I_derivation_4} is due to \eqref{con_pdf}, ${d_{n,k}} \buildrel \Delta \over = \frac{{{{\left| {{H_i}\sqrt {{p_i}} \left( {{X_{i,n}} - {X_{i,k}}} \right) + {Z_i}} \right|}^2}}}{{{\sigma ^2}W}}$ and the first term in \eqref{E_zi_2} is based on ${\mathbb{E}_{{Z_i}}}\left\{ {{{\left| {{Z_i}} \right|}^2}} \right\} = {\sigma ^2}W$.
\end{appendices}

\bibliographystyle{IEEE-unsorted}
\bibliography{refs0308}

\end{document}